\documentclass[peerreview,12pt,twocolumn,double,draftclsnofoot,romanappendices]{IEEEtran}
\usepackage[cmex10]{amsmath}
\usepackage{graphics}
\usepackage{mdwmath,mdwtab,url}
\usepackage[caption=false,font=footnotesize]{subfig}
\usepackage{stfloats}
\usepackage{nomencl}
\usepackage{amssymb}
\usepackage{rotating}
\usepackage{psfrag}
\usepackage{color}
\usepackage{hyperref}
\usepackage{cite}
\usepackage{epstopdf}

\newtheorem{theorem}{\textbf{Theorem}}
\newtheorem{corollary}{\textbf{Corollary}}

\newtheorem{definition}{{Definition}}

\newtheorem{remark}{\textbf{Remark}}
\newcommand{\mc}{\mathcal} 

\newcommand{\tc}{\textcolor}
\definecolor{orange}{rgb}{1,0.5,0.0}
\begin{document}

\title{Secure Communication over Parallel Relay Channel}

\author{Zohaib~Hassan~Awan, Abdellatif~Zaidi, and Luc~Vandendorpe,~\IEEEmembership{Fellow,~IEEE} 
				
\thanks{Copyright (c) 2010 IEEE. Personal use of this material is permitted. However, permission to use this material for any other purposes must be obtained from the IEEE by sending a request to pubs-permissions@ieee.org.}
				
\thanks{Zohaib Hassan Awan and Luc Vandendorpe are with ICTEAM institute (\'{E}cole Polytechnique de Louvain), Universit\'e catholique de Louvain, Louvain-la-Neuve 1348, Belgium. Email: \{zohaib.awan, luc.vandendorpe\}@uclouvain.be}

\thanks{Abdellatif Zaidi was with the ICTEAM institute (\'{E}cole Polytechnique de Louvain), Universit\'e catholique de Louvain, Louvain-la-Neuve 1348, Belgium, and is now with Universit\'e Paris-Est Marne-la-Vall\'ee, 77454 Marne-la-Vall\'ee Cedex 2, France. Email: abdellatif.zaidi@univ-mlv.fr} 

\thanks{This work was supported in part by the EU network of excellence NEWCOM++, and the Concerted Research Action, SCOOP.}

\thanks{ The result in this work was presented in part at 48th Annual Allerton Conference on Communication, Control, and Computing, Monticello, IL, USA, Sept. 2010.}}

\markboth{To appear in IEEE Transactions on Information Forensics and Security }{Secure Communication over Parallel Relay Channel}

\maketitle

\begin{abstract}
We investigate the problem of secure communication over parallel relay channel in the presence of a passive eavesdropper. We consider a four terminal relay-eavesdropper channel which consists of multiple relay-eavesdropper channels as subchannels. For the discrete memoryless model, we establish outer and inner bounds on the rate-equivocation region. The inner bound allows mode selection at the relay. For each subchannel, secure transmission is obtained through one of two coding schemes at the relay: decoding-and-forwarding the source message or confusing the eavesdropper through noise injection. For the Gaussian memoryless channel, we establish lower and upper bounds on the perfect secrecy rate. Furthermore, we study a special case in which the relay does not hear the source and show that under certain conditions the lower and upper bounds coincide. The results established for the parallel Gaussian relay-eavesdropper channel are then applied to study the fading relay-eavesdropper channel. Analytical results are illustrated through some numerical examples.
\end{abstract}


\newpage
\section{Introduction}
In conventional point-to-point wired networks, security is facilitated by secret key sharing between relevant parties based on some common cryptographic algorithm. The premise is that only legitimate users have access to the encrypted messages and extraneous users (adversaries) are unable to access any useful information. The wireless channel is characterized by its inherit randomness and broadcast nature. Physical layer security exploits the basic attributes of the wireless channel for instance, difference of the fading gains between the legitimate channel (source to the legitimate receiver) and the channel to the adversary, to transmit information securely to the legitimate receiver. Thus, it eradicates the need of secret key sharing. 

The wiretap channel introduced by Wyner is a basic information-theoretic model which incorporates physical layer attributes of the channel to transmit information securely \cite{6}.  Wyner's basic model consists of a source, a legitimate receiver and an eavesdropper (wiretapper) under noisy channel conditions. Secrecy capacity is established when the eavesdropper channel (the channel from the source to the eavesdropper) is a degraded version of the main channel (the channel from the source to the legitimate receiver). The discrete memoryless (DM) channel studied by Wyner is further extended to study some other channels for which secrecy capacity is established, i.e., broadcast channels (BC) \cite{9,15}, multi-antenna channels \cite{17,19,21}, multiple access channels \cite{27,31,28}, fading channels \cite{29,30} etc. The idea of cooperation between users in context of security was introduced by \cite{10}. The intuition is that, when the main channel is more noisy than the channel to the eavesdropper, cooperation between users is utilized to achieve positive secrecy capacity. Secrecy is achieved by using the relay as a trusted node that facilitates the information decoding at the destination while concealing the information from the eavesdropper. A special case in which there is a physically degraded relay-eavesdropper channel was studied in \cite{25}. The case in which the relay does not acts as a trusted node is studied in \cite{26,32}.

In this paper, we study a parallel relay-eavesdropper channel. A parallel relay-eavesdropper channel is a generalization of the setup in \cite{10}, in which each of the source-to-relay (S-R), source-to-destination (S-D), source-to-eavesdropper (S-E), relay-to-destination (R-D) and relay-to-eavesdropper (R-E) link is composed of several parallel channels as subchannels. The eavesdropper is passive in the sense that it just listens to the transmitted information without modifying it. We only focus on the \textit{perfect secrecy rate}, i.e., the maximum achievable rate at which information is reliably sent to the legitimate receiver, and the eavesdropper is unable to decode it. 

The parallel relay-eavesdropper channel considered in this paper relates to some of the channels studied previously. Compared to the parallel relay channel studied in \cite{13}, the parallel relay-eavesdropper channel requires an additional secrecy constraint. The parallel relay-eavesdropper channel without relay simplifies to a number of channels discussed previously. For example, the parallel wiretap channel studied in \cite{23}, the parallel broadcast channel with confidential messages (BCC) and no common message studied in \cite{15}.

\vspace{.5em}
\textit{\textbf{Contributions.}} The main contributions of this paper are summarized as follows. For the discrete memoryless case, we establish inner and outer bounds on the rate-equivocation region for the parallel relay-eavesdropper channel. The inner bound is obtained through a coding scheme in which, for each subchannel, the relay operates either in decode-and-forward (DF) or in noise forwarding (NF) mode. We note that establishing our outer bound for DM case is not straightforward and it does not follow directly from the single-letter outer bound for the relay-eavesdropper channel developed in \cite[Theorem 1]{10}. Therefore a converse is needed. The converse includes a re-definition of the involved auxiliary random variables, a technique much similar to the one used before in the context of secure transmission over broadcast channels \cite{15}.

For the Gaussian memoryless model, we establish lower and upper bounds on the perfect secrecy rate. The lower bound established for the Gaussian model follows directly from the DM case. However, we note that establishing a computable upper bound on the secrecy rate for the Gaussian model is non-trivial, and it does not follow directly from the DM case. In part, this is because the upper bound established for the DM case involves auxiliary random variables, the optimal choice of which is difficult to obtain. In this work, we develop a new upper bound on the secrecy rate for the parallel Gaussian relay-eavesdropper channel. Our converse proof uses elements from converse techniques developed in \cite{19,21} in context of multi-antenna wiretap channel; and in a sense, can be viewed as an extension of these results to the parallel relay-eavesdropper channel. This upper bound is especially useful when the multiple access part of the channel is the bottleneck. We show that, in contrast to upper bounding techniques for our model that
can be obtained straightforwardly by applying recent results on multi-antenna wiretap channels \cite{17,19,21}, our upper bound shows some degree of \textit{separability} for the different subchannels.

We also study a special case in which the relay does not hear the source, for example due to very noisy source-to-relay links. In this case we show that under some specific conditions noise-forwarding on all links achieves the secrecy capacity. The converse proof follows from 	a new genie-aided upper bound that assumes full cooperation between the relay and the destination, and a constrained eavesdropper. The eavesdropper is constrained in the sense that it has to treat the relay's transmission as unknown noise for all subchannels, an idea used previously in context of a class of classic relay-eavesdropper channel with orthogonal components \cite{16}. These assumptions turn the parallel Gaussian relay-eavesdropper channel into a parallel Gaussian wiretap channel, the secrecy capacity of which is established in \cite{15,23}.

Furthermore, we study an application of the results established for the parallel Gaussian relay-eavesdropper channel to the fading relay-eavesdropper channel. We assume that perfect non-causal channel state information (CSI) is available at all nodes. The fading relay-eavesdropper channel is a special case of the parallel Gaussian relay-eavesdropper channel in which each realization of a fading state corresponds to one subchannel. We illustrate our results through some numerical examples.

The rest of the paper is organized as follows. In section II, we establish outer and inner bounds on the rate-equivocation region for the DM channel. In section III, we establish lower and upper bounds on the perfect secrecy rate for the Gaussian model, and consider a special case in which under some specific conditions secrecy capacity is achieved. In section IV, we present an application of the results established in section III to the fading model. We illustrate these results with some numerical examples in section V. Section VI concludes the paper by summarizing its contribution.

\vspace{.5em}
\textit{\textbf{Notations.}}  In this paper, the notation $X_{[1,L]}$ is used as a shorthand for $(X_1,X_2,\hdots,X_L)$, the notation $X_{[1,L]}^n$ is used as a shorthand for  $(X_1^n,X_2^n,\hdots,X_L^n)$ where for $l=1,\hdots,L$,  $X_l^n:=(X_{l1},X_{l2},\hdots,X_{ln})$, the notation $X_{[1,L],i}$ is used as a shorthand for $(X_{1,i},X_{2,i},\hdots,X_{L,i})$, the notation $\mc X_{1[1,L]}$ is used as a shorthand for $\mc X_{11}\times \mc X_{12}\hdots\times \mc X_{1L}$, $\mathbb{E}\{.\}$ denotes the expectation operator, $|\mc{X}|$ denotes the cardinality of set $\mc{X}$, $L$ denotes the number of subchannels, the boldface letter $\bf{X}$ denotes the covariance matrix. We denote the entropy of a discrete and continuous random variable $X$ by $H(X)$ and $h(X)$ respectively. We define the functions $\mc{C}(x)=\frac{1}{2}\log_2(1+x)$ and $[x]^+ = \max\{0,x\}$. Throughout the paper the logarithm function is taken to the base 2.

\section{Discrete memoryless channel}
In this section, we establish outer and inner bounds on the rate-equivocation region for the discrete memoryless parallel relay-eavesdropper channel. 
\begin{figure}
\psfragscanon
\begin{center}
\psfrag{W}[c][c]{$W$}
\psfrag{Q}[c][c]{$Y_{1}^n$}
\psfrag{P}[c][c]{$Y_{L}^n$}
\psfrag{Z}[c][c]{\tc{red}{$Y_{21}^n$}}
\psfrag{S}[c][c]{\tc{red}{$Y_{2L}^n$}}
\psfrag{Y}[c][c]{{$Y_{11}^n$}}
\psfrag{K}[c][c]{{$Y_{1L}^n$}}
\psfrag{R}[c][c]{{Relay}}
\psfrag{E}[c][c]{{Encoder}}
\psfrag{V}[c][c]{$X_{21}^n$}
\psfrag{B}[c][c]{$X_{2L}^n$}
\psfrag{X}[c][c]{\hspace{-.1cm}$X_{11}^n$}
\psfrag{G}[c][c]{\hspace{-.1cm}$X_{1L}^n$}
\psfrag{O}[c][c]{\:\:$\hat{W}$}
\psfrag{C}[c][c]{$p(y_{[1,L]},y_{1[1,L]},y_{2[1,L]}|x_{1[1,L]},x_{2[1,L]})$}
\psfrag{D}[c][c]{{Decoder}}
\psfrag{F}[c][c]{{{Eavesdropper}}}
\includegraphics[width=\linewidth]{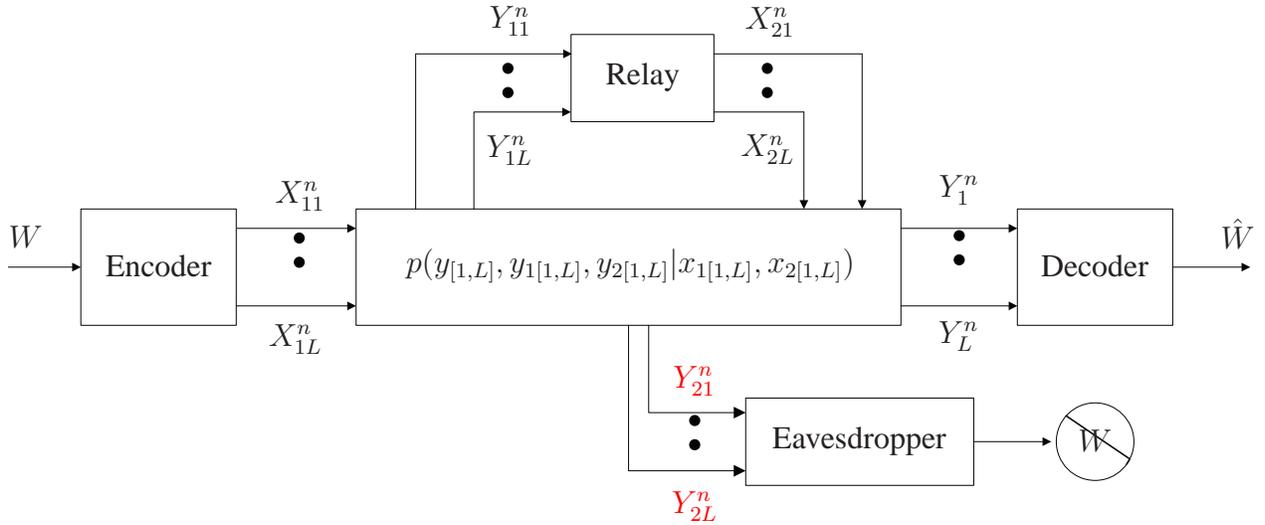}
\end{center}
\caption{The parallel relay-eavesdropper channel.}
\psfragscanoff
 \label{dmc}
\end{figure}

\subsection{Channel Model}
\begin{definition}
The parallel relay-eavesdropper channel consists of four nodes, a source, a relay, a destination (legitimate receiver) and a passive eavesdropper. The communication takes place over $L$ subchannels. Fig. 1 represents the studied model. The source wishes to send confidential messages to the destination, with the help of the relay to conceal them from passive eavesdropper. The source encodes the confidential message $W$ to $(X_{11}^n,X_{12}^n,\hdots,X_{1L}^n)$ codewords and broadcasts it over $L$ subchannels to the relay and the destination. The relay helps to reduce the uncertainty about the confidential message at the destination by re-encoding whatever it has received from the source and transmits $(X_{21}^n,X_{22}^n,\hdots,X_{2L}^n)$ codewords to the destination.
The outputs at the relay and destination are given by $(Y_{11}^n,Y_{12}^n,\hdots,Y_{1L}^n)$ and $(Y_{1}^n,Y_{2}^n,\hdots,Y_{L}^n)$ respectively. The passive eavesdropper overhears to the source and relay transmission over the multiple-access link, which is denoted by $(Y_{21}^n,Y_{22}^n,\hdots,Y_{2L}^n)$. 

More precisely, the parallel relay-eavesdropper channel consists of  $\mc X_{1[1,L]}, \mc X_{2[1,L]}$ as finite input alphabets and $\mc Y_{[1,L]}, \mc Y_{1[1,L]}, \mc Y_{2[1,L]}$ as finite output alphabets. Since the channel is memoryless, the transition probability distribution is given by
\begin{eqnarray}
\label{eq1}
\prod_{l=1}^L\prod_{i=1}^{n} p(y_{l,i},y_{1l,i},y_{2l,i} \mid x_{1l,i},x_{2l,i})
\end{eqnarray}
where $x_{1l,i} \in \mc X_{1l}, x_{2l,i} \in \mc X_{2l},  y_{l,i} \in \mc Y_{l}, y_{1l,i} \in \mc Y_{1l}$ and $y_{2l,i} \in \mc Y_{2l}$,  for $l= 1,\hdots,L$ and $i= 1,\hdots,n$. The symbols $x_{1l}$ and $x_{2l}$ are the source and relay inputs on subchannel $l$, and $ y_{1l}, y_{l}, y_{2l}$ are the channel outputs at the relay, destination and eavesdropper for the $l$-th subchannel respectively.
\end{definition}
\vspace{0.5em}
\begin{definition}
The source sends a message ${W} \in \mathcal{W} = \{1,\hdots,2^{n{{R}}} \}$ using a ($2^{nR},n$) code consisting of
\begin{itemize}
\item a stochastic encoder at the source that maps $W \rightarrow {X}_{1[1,L]}^n$,
\item a relay encoder that maps $f_i(Y_{1[1,L]}^{i-1})\rightarrow X_{2[1,L],i}$ for $1 \le i \le n$,
\item a decoding function $g(.)$, that maps the received codewords from the source and relay node to get an estimate of the confidential message, $g(Y_{[1,L]}^n) \rightarrow \hat{W}$.
\end{itemize}
\end{definition}
\vspace{0.5em}
\begin{definition}
The average error probability of a ($2^{nR},n$) code is defined as
\begin{eqnarray}
P_{e}^n = \frac{1}{2^{n{R}}} \sum_{W \in \mathcal{W}}p\{ g(Y_{[1,L]}^n) \ne W | W  \}.
\end{eqnarray}
\end{definition}
\vspace{0.5em}
Due to the openness of the wireless medium, the eavesdropper listens for free to what the source and relay transmit. It then tries to guess the information being transmitted. The equivocation rate per channel use is defined as $R_e = H(W|Y_{2[1,L]}^n)/n$. Perfect secrecy for the channel is obtained when the eavesdropper gets no information about the confidential message $W$ from $ Y_{2[1,L]}^n$. That is, the equivocation rate is equal to the unconditional source entropy.
\vspace{0.5em}
\begin{definition}[\cite{6}]
A rate-equivocation pair (${{R},{R}_{e}}$) is achievable for the parallel relay-eavesdropper channel, if for any $\epsilon > 0$ there exists a sequence of codes ($2^{n{R}}$, $n$) such that for any $n \ge n(\epsilon)$ 
\setlength{\arraycolsep}{0.2em}
\begin{eqnarray}
\label{def4}
\frac{H(W)}{n} &\ge& R - \epsilon,\notag \\
\frac{H(W| Y_{2[1,L]}^n)}{n} &\ge& R_e - \epsilon , \notag \\
P_{e}^n &\le& \epsilon.
\end{eqnarray}
\end{definition}
\setlength{\arraycolsep}{5pt}
\subsection{Outer Bound}
The following theorem provides an outer bound on the rate-equivocation region for the parallel relay-eavesdropper channel.
\vspace{0.5em}
\begin{theorem}
\label{Up}
For a parallel relay-eavesdropper channel with $L$ subchannels, and for any achievable rate-equivocation pair $(R,R_e)$, there exists a set of random variables ${U_l}\rightarrow({V_{1l},V_{2l}})\rightarrow({X_{1l},X_{2l}})\rightarrow({Y_l,Y_{1l},Y_{2l}})$, $l=1,\hdots,L$, such that ($R,R_e$) satisfies
\setlength{\arraycolsep}{0.2em}
\begin{eqnarray}
\label{Upeq}
{R}   &\le& \min \big \{ \sum_{l=1}^L {I({V_{1l},V_{2l}};Y_l)},\sum_{l=1}^{L} {I(V_{1l};Y_l,Y_{1l} \mid V_{2l})} \big\} \notag \\
{R}_e &\le& {R} \notag \\
{R}_e &\le& \min \big\{\sum_{l=1}^L I(V_{1l},V_{2l};Y_l\mid U_l)-{I(V_{1l},V_{2l};Y_{2l}\mid U_l)},\notag\\&&\quad \sum_{l=1}^L {I(V_{1l};Y_l,Y_{1l} \mid V_{2l},U_l)}-{I(V_{1l},V_{2l};Y_{2l}\mid U_l)}\big\}.
 \end{eqnarray}
 \setlength{\arraycolsep}{5pt}
\end{theorem}
\vspace{0.5em}
\begin{IEEEproof}
The proof of Theorem 1 is given in Appendix \ref{app1}.
\end{IEEEproof}

\vspace{0.5em}
\begin{remark}
The outer bound in Theorem 1 does not follow directly from the single-letter outer bound on the rate-equivocation region established for the relay-eavesdropper channel \cite[Theorem 1]{10}. Therefore a converse is required, in which we need to re-define the involved auxiliary random variables. The technique used to re-define the auxiliary random variables has some connection with the one used before in the context of secure transmission over broadcast channels \cite{15}.           
\end{remark}

\vspace{0.5em}
\begin{remark}
The region \eqref{Upeq} reduces to the rate-equivocation region developed for the relay-eavesdropper channel \cite[Theorem 1]{10} by setting $L:=1$ in \eqref{Upeq}.
\end{remark}

\vspace{0.5em}
\begin{remark}
The equivocation rate in Theorem 1 reduces to the secrecy capacity of the parallel wiretap channel established in \cite[Corollary 1]{15} by removing the relay, i.e., by setting $Y_{1l}=X_{2l}=V_{2l}=\phi$. The resulting term $\sum_{l=1}^L I(V_{1l};Y_l\mid U_l)-{I(V_{1l};Y_{2l}\mid U_l)}$ is maximized by $U_l:$=constant, for $l=1,\hdots,L$.
\end{remark}
\vspace{0.5em}

\subsection{Achievable Rate-Equivocation Region}
In this subsection we establish an achievable rate-equivocation region for the parallel relay-eavesdropper channel. The achievable region is established by the combination of two different coding schemes, namely decode-and-forward and noise forwarding. In DF scheme,  for each message source associates a number of confusion codewords, the relay after receiving the source codewords, decode it and re-transmits it towards the legitimate receiver and eavesdropper (see \cite[Theorem 2]{10} for details). In the NF scheme the relay does not decode the source codewords, but transmits  confusion codewords independent from the source codewords, towards the legitimate receiver and the eavesdropper (see \cite[Theorem 3]{10} for details).

\vspace{0.5em}
\begin{theorem}
\label{low}
For a parallel relay-eavesdropper channel with $L$ subchannels, the rate pairs in the closure of the convex hull of all ($R, R_{e}$) satisfying
\setlength{\arraycolsep}{0.2em}
\begin{eqnarray}
\label{innerd}
R &\le&  \min \big \{\sum_{l\in \mc{A}}I(V_{1l},V_{2l};Y_l|U_l),\sum_{l\in \mc{A}} I(V_{1l};Y_{1l} | V_{2l},U_l)\big\} +\sum_{l\in \mc{A}^c} I(V_{1l};Y_l|V_{2l}) \notag \\
R_e &\le& R \notag \\
R_e &\le&  \min \big \{\sum_{l\in \mc{A}}  I(V_{1l},V_{2l};Y_l| U_l) -I(V_{1l},V_{2l};Y_{2l}| U_l),\sum_{l\in \mc{A}} I(V_{1l};Y_{1l} | V_{2l},U_l)-I(V_{1l},V_{2l};Y_{2l}| U_l) \big\} \notag \\
&&+\sum_{l\in \mc{A}^c} I(V_{1l};Y_l | V_{2l})+ \min \big\{\sum_{l\in \mc{A}^c}I(V_{2l};Y_l),\sum_{l\in \mc{A}^c} I(V_{2l};Y_{2l} | V_{1l})\big\}-\min \big \{\sum_{l\in \mc{A}^c}I(V_{2l};Y_l),\notag\\&& \sum_{l\in \mc{A}^c}I(V_{2l};Y_{2l})\big\}-\sum_{l\in \mc{A}^c} I(V_{1l};Y_{2l}| V_{2l}),
\end{eqnarray}
\end{theorem}
\setlength{\arraycolsep}{5pt}
\noindent for some distribution \small{$p(u_l,v_{1l},v_{2l},x_{1l},x_{2l},y_{l},y_{1l},y_{2l})=p(u_l)p(v_{1l},v_{2l}|u_l)p(x_{1l},x_{2l}| v_{1l},v_{2l})p(y_{l},y_{1l},y_{2l}|x_{1l},x_{2l})$} \normalsize for $l\in {\mc{A}}$ and \small{$p(v_{1l},v_{2l},x_{1l},x_{2l},y_{l},y_{1l},y_{2l}) = p(v_{1l})p(v_{2l})p(x_{1l}|v_{1l})p(x_{2l}|v_{2l})p(y_{l},y_{1l},y_{2l}|x_{1l},x_{2l})$} \normalsize for $l \in {\mc{A}^c}$, are achievable.

\vspace{.5em}
\begin{IEEEproof}[Outline of Proof] \\
 The region in Theorem \ref{low} is obtained through a coding scheme which combines appropriately DF and NF schemes. In the statement of Theorem \ref{low}, sets $\mc{A}$ and $\mc{A}^c$ represent the subchannels for which relay operates in DF and NF mode, respectively. The rates for the DF scheme can be obtained readily by setting $U := U_{[1,|\mc {A}|]},V_{1}:= V_{1[1,|\mc {A}|]}, V_2 := V_{2[1,|\mc {A}|]},Y := Y_{[1,|\mc {A}|]}$, $Y_1 := Y_{1[1,|\mc {A}|]}$ and $Y_2 := Y_{2[1,|\mc {A}|]}$, for $l \in \mc{A}$ in \cite[Theorem 2]{10}. Similarly the rates for NF scheme can be readily obtained by setting $V_{1}:= V_{1[1,|\mc{A}^c|]}, V_2 := V_{2[1,|\mc{A}^c|]}, Y := Y_{[1,|\mc{A}^c|]}$, $Y_1 := Y_{1[1,|\mc{A}^c|]}$ and $Y_2 := Y_{2[1,|\mc{A}^c|]}$, for $l \in \mc{A}^c$ in \cite[Theorem 3]{10}.
\end{IEEEproof}

\vspace{0.5em}
\begin{remark}
For a parallel relay-eavesdropper channel in which all subchannels are degraded{\footnote{In parallel relay-eavesdropper channel if all subchannels  are degraded, the entire relay-eavesdropper channel may not necessarily be degraded.}}, i.e.,
\begin{eqnarray}
 p(y_l,y_{1l},y_{2l} | x_{1l},x_{2l}) =  p(y_{1l}| x_{1l},x_{2l})p(y_{l}| y_{1l},x_{2l})p(y_{2l}|y_{l}, y_{1l}, x_{1l},x_{2l}),\notag
\end{eqnarray}
 $l=1,\hdots,L$, the perfect secrecy capacity is given by
\setlength{\arraycolsep}{0.2em}
\begin{align}
\label{pd}
C_s =& \max \min \big\{\sum_{l=1}^L [I(V_{1l},V_{2l};Y_l\mid U_l)-{I(V_{1l},V_{2l};Y_{2l}\mid U_l)}]^+, \notag \\&\quad\quad\quad\quad\sum_{l=1}^L [{I(V_{1l};Y_{1l} \mid V_{2l},U_l)}-{I(V_{1l},V_{2l};Y_{2l}\mid U_l)}]^+\big\}
 \end{align}
where the maximization is over ${U_l}\rightarrow({V_{1l},V_{2l}})\rightarrow({X_{1l},X_{2l}})\rightarrow({Y_l,Y_{1l},Y_{2l}})$, for $l=1,\hdots,L$.
 
\vspace{0.5em}
\begin{IEEEproof}
 The achievability follows from Theorem 2 by setting $\mc A^c:=\varnothing$. The converse follows along the lines of Theorem \ref{Up} and is omitted for brevity.
\end{IEEEproof}
\end{remark}



\section{Gaussian Channel}
\begin{figure}[h]
\centering
\includegraphics[width=\linewidth]{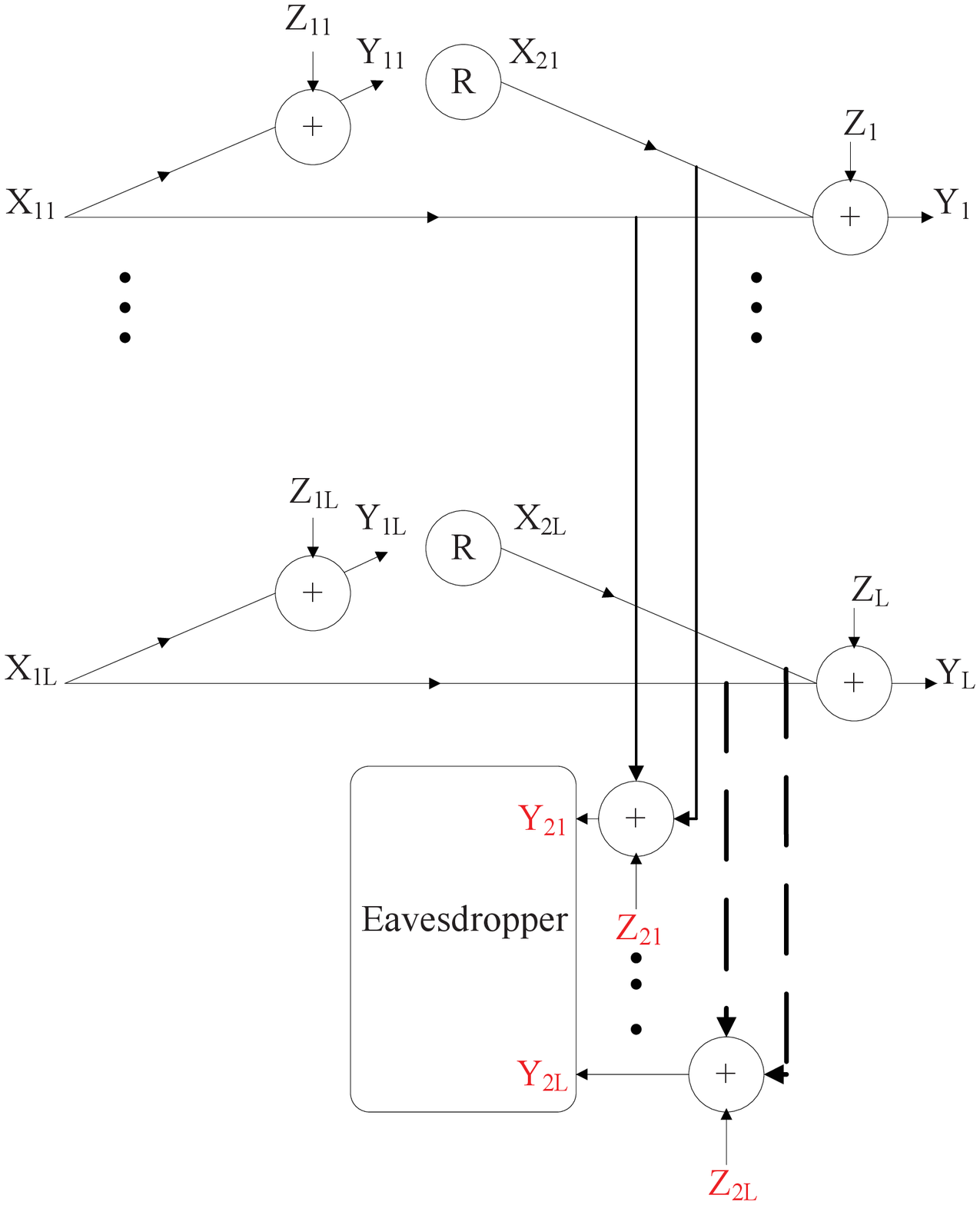}
\caption{The parallel Gaussian relay-eavesdropper channel.}
 \label{fig2}
\end{figure}

In this section we study a parallel Gaussian relay-eavesdropper channel. Fig. \ref{fig2} depicts the studied model. We only focus on the perfectly secure achievable rates, i.e., $(R,R_e)= (R,R)$. 

\subsection{Channel Model}
For a parallel Gaussian relay-eavesdropper channel, the received signals at the relay, destination and eavesdropper are given by%
\begin{align}
\label{gchan}
{Y_{1l,i}}&= {X_{1l,i}}+{Z_{1l,i}}\notag\\
{Y_{l,i}} &= {X_{1l,i}}+\sqrt{\rho_{1l}}{X_{2l,i}}+{Z_{l,i}}\notag\\
{Y_{2l,i}}&= {X_{1l,i}}+\sqrt{\rho_{2l}}{X_{2l,i}}+{Z_{2l,i}}
\end{align}
where $i$ is the time index, $\{Z_{1l,i}\},\{Z_{l,i}\}$ and $\{Z_{2l,i}\}$ are noise processes, independent and identically distributed (i.i.d) with the components being zero mean Gaussian random variables with variances $\sigma_{1l}^2$, $\sigma_{l}^2$ and $\sigma_{2l}^2$ respectively, for $l=1,\hdots,L$. We assume that the source and relay know the noise variances present at the receivers. For the subchannel $l$, $X_{1l,i}$ and $X_{2l,i}$ are inputs from the source and relay nodes respectively. The parameter $\rho_{1l}$ indicates the ratio of the R-D link signal-to-noise (SNR) to the S-D link SNR and $\rho_{2l}$ indicates the ratio of the R-E link SNR to the S-E link SNR for subchannel $l$ respectively. The source and relay input sequences are subject to separate power constraints $P_1$ and $P_2$, i.e.,
\begin{eqnarray}
\label{p_con1}
\frac{1}{n} \sum_{l=1}^L\sum_{i=1}^n \mathbb{E}[X_{1l,i}^2]\le  P_1, \\
 \label{p_con2}
 \frac{1}{n}  \sum_{l=1}^L \sum_{i=1}^n \mathbb{E}[X_{2l,i}^2]\le P_2.
\end{eqnarray}


\begin{figure*}[t]
\centering
\includegraphics[width=\linewidth]{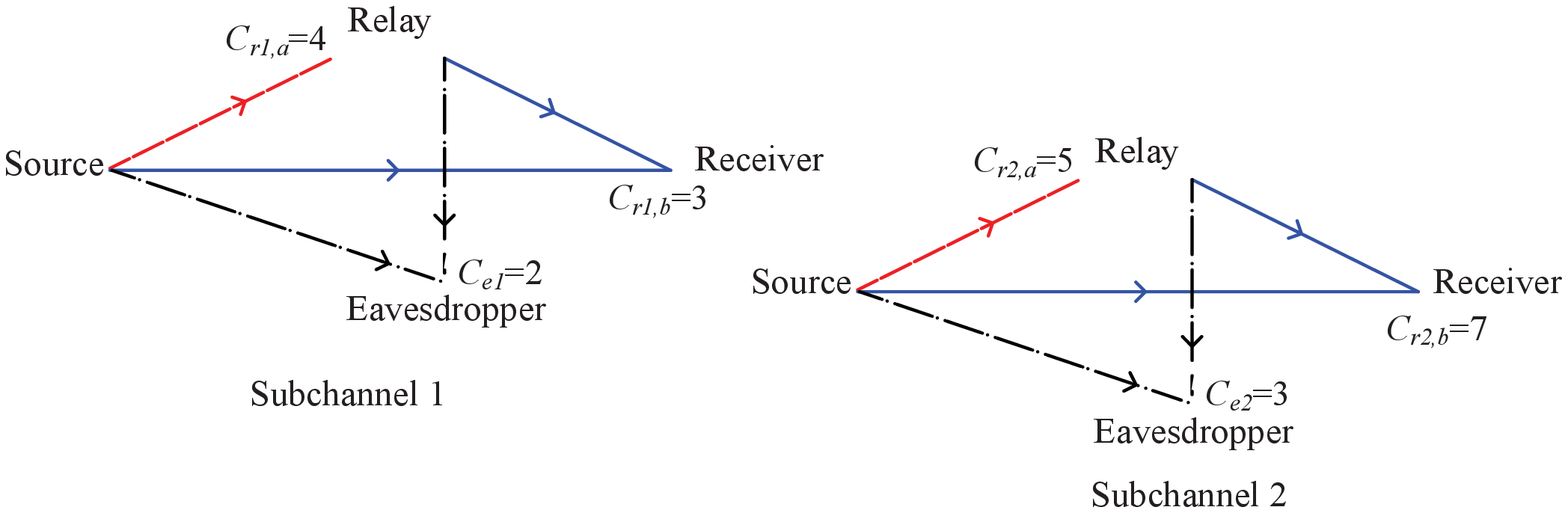}
\caption{An example of a deterministic parallel relay-eavesdropper channel with two subchannels.}
 \label{remark}
\end{figure*}

\subsection{Lower Bound on the Perfect Secrecy Rate}
For the parallel Gaussian relay-eavesdropper channel \eqref{gchan}, we apply Theorem \ref{low} to obtain a lower bound on the perfect secrecy rate.\footnote{The results established for the DM case can be readily extended to memoryless channels with discrete time and continuous alphabets using standard techniques \cite[Chapter 7]{22}.}
\vspace{0.5em}
\begin{corollary}
\label{cor1}
For the  parallel Gaussian relay-eavesdropper channel (\ref{gchan}), a lower bound on the perfect secrecy rate is given by
{\small{
\begin{align}
\label{glow}
R_{e}^{\textrm{low}} =&
{\max_{
\substack{
\sum_{l=1}^L P_{1l} \le P_1, \sum_{l=1}^L P_{2l} \le P_2, \\
0 \le\alpha_{l}\le 1,\hspace{.2em} \text{for $l=1,\hdots,|\mc{A}|$}}
}}
 \min \Big \{ \sum_{l\in \mc{A}} \mathcal{C}\Big(
\frac{P_{1l}+\rho_{1l} P_{2l}+2\sqrt{\bar{\alpha}_l\rho_{1l}{P_{1l}P_{2l}}}}{{\sigma}_{l}^2}\Big)\notag\\&-\mathcal{C}\Big(
\frac{{P_{1l}+ \rho_{2l}P_{2l}}+2\sqrt{\bar{\alpha}_l\rho_{2l}{P_{1l}P_{2l}}}}{{\sigma}_{2l}^2}\Big), \sum_{l\in \mc{A}}\mathcal{C} \Big( \frac{\alpha_l{P_{1l}}}{\sigma_{1l}^2} \Big)-\mathcal{C}\Big(
\frac{{P_{1l}+ \rho_{2l}P_{2l}}+2\sqrt{\bar{\alpha}_l\rho_{2l}{P_{1l}P_{2l}}}}{{\sigma}_{2l}^2}\Big) \Big \}\notag\\&
+\sum_{l \in \mc{A}^c} \mathcal{C} \Big( \frac{{P_{1l}}}{{\sigma}_{l}^2}\Big)+ \min \Big \{ \sum_{l \in \mc{A}^c}\mathcal{C}\Big(\frac{\rho_{1l}P_{2l}}{P_{1l}+{\sigma}_{l}^2}\Big), \sum_{l \in \mc{A}^c}\mathcal{C}\Big(\frac{\rho_{2l}P_{2l}}{{\sigma}_{2l}^2}\Big) \Big\}
\notag\\&- \min \Big \{\sum_{l \in \mc{A}^c} \mathcal{C}\Big(\frac{\rho_{1l}P_{2l}}{P_{1l}
+{\sigma}_{l}^2}\Big),\sum_{l \in \mc{A}^c} \mathcal{C}\Big(\frac{\rho_{2l}P_{2l}}{P_{1l}+{\sigma}_{2l}^2}\Big) \Big\}- \sum_{l \in \mc{A}^c}\mathcal{C}\Big(\frac{P_{1l}}{{\sigma}_{2l}^2}\Big).
\end{align}}}
\end{corollary}
\vspace{0.5em}
\begin{IEEEproof}
 The achievability follows by applying Theorem \ref{low} with the choice $U_l:=$ constant, $V_{1l}:=X_{1l}$, $V_{2l}:= X_{2l}$, ${X_{1l}}:={\tilde{X}_{1l}} + \sqrt{\frac{\bar{\alpha}_l P_{1l}}{P_{2l}}} {X_{2l}}$, $\bar{\alpha}_l:=1-\alpha_l$, $\tilde{X}_{1l}\sim\mc N(0,\alpha_lP_{1l})$ independent of $ X_{2l}\sim\mc N(0,P_{2l})$, where $\alpha_l \in [0,1]$ for $l \in {\mc{A}}$; and $X_{1l}\sim\mc N(0,P_{1l})$ independent of $X_{2l}\sim\mc N(0,P_{2l})$ for $l\in\mc{A}^c$. Straightforward algebra which is omitted for brevity gives \eqref{glow}.
\end{IEEEproof}
\vspace{0.5em}
The parameters $P_{1l}$ and $P_{2l}$ indicate the source and relay power allocated for transmission over the $l$-th subchannel. In \eqref{glow},  after some straightforward algebra, the contribution to the equivocation of information sent through NF (set $\mc{A}^c$ in Theorem \ref{low}) can be condensed by observing that we only need to consider $\min\{\sum_{l \in \mc{A}^c}I(X_{2l};Y_{2l}),\sum_{l \in \mc{A}^c}I(X_{2l};Y_{l})\}=\sum_{l \in \mc{A}^c}I(X_{2l};Y_{2l})$, to get higher secrecy rate. A simplified expression for $R_e^{\textrm{low}}$ is given by 
{\small{
\begin{align}
\label{glowm}
{R}_{e}^{\textrm{low}} =&
\max_{
\substack{
\sum_{l=1}^L P_{1l} \le P_1, \sum_{l=1}^L P_{2l} \le P_2, \\
0 \le\alpha_{l}\le 1,\hspace{.2em} \text{for $l=1,\hdots,|\mc{A}|$}}
}
 \min \Big \{ \sum_{l\in \mc{A}} \Big[\mathcal{C}\Big(
\frac{P_{1l}+\rho_{1l} P_{2l}+2\sqrt{\bar{\alpha}_l\rho_{1l}{P_{1l}P_{2l}}}}{{\sigma}_{l}^2}\Big)\notag\\& -\mathcal{C}\Big(
\frac{{P_{1l}+ \rho_{2l}P_{2l}}+2\sqrt{\bar{\alpha}_l\rho_{2l}{P_{1l}P_{2l}}}}{{\sigma}_{2l}^2}\Big)\Big]^+ ,\sum_{l\in \mc{A}}\Big[\mathcal{C} \Big( \frac{\alpha_l{P_{1l}}}{\sigma_{1l}^2} \Big)-\mathcal{C}\Big(
\frac{{P_{1l}+ \rho_{2l}P_{2l}}+2\sqrt{\bar{\alpha}_l\rho_{2l}{P_{1l}P_{2l}}}}{{\sigma}_{2l}^2}\Big)\Big]^+ \Big \}\notag\\&
+ \min\Big \{\sum_{l \in \mc{A}^c}\Big[\mathcal{C} \Big( \frac{P_{1l}+\rho_{1l}P_{2l}}{{\sigma}_{l}^2}\Big)-\mathcal{C}\Big(\frac{P_{1l}+\rho_{2l}P_{2l}}{{\sigma}_{2l}^2}\Big)\Big]^+, \sum_{l \in \mc{A}^c}\Big[\mathcal{C}\Big(\frac{P_{1l}}{{\sigma}_{l}^2}\Big)+\mathcal{C}\Big(\frac{\rho_{2l}P_{2l}}{{\sigma}_{2l}^2}\Big)\notag\\&-\mathcal{C}\Big(\frac{P_{1l}+\rho_{2l}P_{2l}}{{\sigma}_{2l}^2}\Big)\Big]^+\Big\}.
\end{align}}}\normalsize{\noindent In \eqref{glowm}, for each subchannel $[.]^+$ appears because achievable secrecy rate is always non-negative.}
\normalsize

\vspace{0.5em}
\begin{remark}
The achievable perfect secrecy rate established in Corollary \ref{cor1} can be larger than the one obtained by coding separately over different parallel subchannels.
\end{remark}
\vspace{0.5em}
This remark is elucidated by the following example. \vspace{.5em}

\textit{Example:} We consider a deterministic parallel relay-eavesdropper channel with two subchannels, i.e., $L:=2$, as shown in Fig. \ref{remark}. For subchannel 1, the link capacities to the relay, legitimate receiver and eavesdropper are given by $C_{r1,a}= 4, C_{r1,b}= 3$ and $C_{e1}= 2$ respectively. For subchannel 2, the link capacities to the relay, legitimate receiver and eavesdropper are given by $C_{r2,a}=5 , C_{r2,b}= 7$ and $C_{e2}= 3$ respectively. For this channel, achievable rate obtained by coding across subchannels is given by

\begin{align}
\label{across}
R_e =& \min\Big\{\sum_{i=1}^2(C_{ri,a}- C_{ei} )^+ ,\sum_{i=1}^2 (C_{ri,b}- C_{ei})^+ \Big\} \notag\\
    =& \min \{4, 5\} = 4.
\end{align}
\noindent Similarly achievable rate obtained by coding separately over each subchannel is given by
\begin{align}
\label{sum}
R_e =& \sum_{i=1}^2\min \{ (C_{ri,a}-C_{ei})^+,(C_{ri,b}-C_{ei})^+ \}    \notag\\
    =& \min \{2, 1\} +\min \{2, 4\} = 3
\end{align}
which is clearly smaller than \eqref{across}. This shows the usefulness of coding across subchannels.
\subsection{Upper Bound on the Perfect Secrecy Rate}
The following theorem provides an upper bound on the secrecy rate for the parallel Gaussian relay-eavesdropper channel.
\vspace{0.5em}
\begin{theorem}
\label{Gup}
For the parallel Gaussian relay-eavesdropper channel \eqref{gchan}, an upper bound on the secrecy rate is given by
\begin{eqnarray}
\label{guu}
{R}_e^{\textrm{up}} =\max_{\{{\bf{K}_{P_\textit{l}}} \in {{\mc{K}_{P_\textit{l}}}}\}_{\textit{l}=1,\hdots, L}}  \sum_{l=1}^L I(X_{1l},X_{2l};Y_l)-I(X_{1l},X_{2l};Y_{2l})
\end{eqnarray}
where the maximization is over $[X_{1l},X_{2l}] \sim \mc{N}(\bf{0},\bf{K}_{P_\textit{l}})$ with ${{\mc{K}}_{P_\textit{l}}}  = \Big \{ {\bf{K}_{P_\textit{l}}} : {\bf{K}_{P_\textit{l}}}$=  $\left [
 \begin{smallmatrix}
  P_{1l} & \psi_\textit{l }\sqrt{P_{1l} P_{2l}}\\
\psi_\textit{l }\sqrt{P_{1l} P_{2l}} & P_{2l}
 \end{smallmatrix} \right ]$, $-1\le \psi_\textit{l} \le {1} \Big\}$,  $l=1,\hdots,L$, with the covariance matrices $\mathbb{E}[X_{1[1,L]}X_{1[1,L]}^T]$,  $\mathbb{E}[X_{2[1,L]}X_{2[1,L]}^T]$ satisfying \eqref{p_con1} and \eqref{p_con2} respectively.
 \setlength{\arraycolsep}{5pt}
\end{theorem}
\vspace{0.5em}
\begin{IEEEproof}
The proof follows from the rate-equivocation region established for the DM case in Theorem \ref{Up}. Taking the first term of minimization in the bound on the equivocation rate in Theorem \ref{Up}, we get
\begin{eqnarray}
\label{nupper}
{R}_e \le  \max \sum_{l=1}^L  I(V_{1l},V_{2l};Y_{l}\mid U_{l})-I(V_{1l},V_{2l};Y_{2l}\mid U_{l})
\end{eqnarray}
where ${U_l}\rightarrow({V_{1l},V_{2l}})\rightarrow({X_{1l},X_{2l}})\rightarrow({Y_l,Y_{1l},Y_{2l}})$, for $l=1,\hdots,L$. The rest of the proof uses elements from related works in \cite{15} and \cite{19}. Continuing from \eqref{nupper}, we obtain
\setlength{\arraycolsep}{.2pt}
\begin{eqnarray}
{R}_e  &\leq& \sum_{l=1}^L  I(V_{1l},V_{2l};Y_{l}\mid U_{l})-I(V_{1l},V_{2l};Y_{2l}\mid U_{l})\notag\\
  &\overset{(a)}{\le}& \sum_{l=1}^L I(V_{1l}, V_{2l};Y_l )-I(V_{1l},V_{2l} ;Y_{2l})\notag \\
  &{\le}& \sum_{l=1}^L I(V_{1l}, V_{2l};Y_l, Y_{2l} )-I(V_{1l},V_{2l} ;Y_{2l})\notag \\
   &\overset{(b)}{=}& \sum_{l=1}^L [I( X_{1l}, X_{2l};Y_l, Y_{2l} )-I( X_{1l}, X_{2l};Y_l, Y_{2l}\mid V_{1l}, V_{2l} )] \notag\\
   &&- [I(X_{1l}, X_{2l} ;Y_{2l})-I(X_{1l}, X_{2l} ;Y_{2l}\mid V_{1l},V_{2l} )]\notag \\
   &{=}& \sum_{l=1}^L [I( X_{1l},X_{2l};Y_l, Y_{2l} )-I(X_{1l}, X_{2l} ;Y_{2l})] \notag\\&
     &- [I( X_{1l}, X_{2l};Y_l,Y_{2l}\mid V_{1l}, V_{2l} )-I(X_{1l}, X_{2l} ;Y_{2l}\mid V_{1l},V_{2l} )]\notag \\
     &{\le}& \sum_{l=1}^L [I( X_{1l}, X_{2l};Y_l, Y_{2l})-I(X_{1l}, X_{2l} ;Y_{2l})] \notag\\
     &=&  \sum_{l=1}^L I( X_{1l}, X_{2l};Y_l \mid Y_{2l})
     \label{ubound}
     \end{eqnarray}
\noindent where (a) follows  by noticing that $I(V_{1l},V_{2l} ;Y_{l}\mid U_{l})-I(V_{1l},V_{2l} ;Y_{2l}\mid U_{l})$ is maximized by $U_l:=\text{constant}$ and (b) follows from the Markov chain condition $({V_{1l},V_{2l}})\rightarrow({X_{1l},X_{2l}})\rightarrow({Y_l,Y_{1l},Y_{2l}})$,  $l =1,\hdots,L$.

We now tighten the upper bound \eqref{ubound} by using an argument previously used in \cite{19}, \cite{21} in the context of multi-antenna wiretap channel. More specifically, observing that, the original bound \eqref{nupper} depends on $p(y_{l},y_{2l}|x_{1l},x_{2l})$ only through its marginals $p(y_{l}|x_{1l},x_{2l})$ and $p(y_{2l}|x_{1l},x_{2l})$, the upper bound \eqref{ubound} can be further tightened as
  \setlength{\arraycolsep}{5pt}
\begin{eqnarray}
\label{nuppern}
{R}_e \le   \min_{\{p(y'_{l},y'_{2l}|x_{1l},x_{2l})\}} \max_{\{p(x_{1l},x_{2l})\}} \sum_{l=1}^L  I( X_{1l}, X_{2l};Y'_l \mid Y'_{2l})
\end{eqnarray}
where the joint conditional $p(y'_{l},y'_{2l}|x_{1l},x_{2l})$ has the same marginals  as $p(y_{l},y_{2l}|x_{1l},x_{2l})$, i.e.,  $p(y'_{l}|x_{1l},x_{2l})=p(y_{l}|x_{1l},x_{2l})$ and $p(y'_{2l}|x_{1l},x_{2l})=p(y_{2l}|x_{1l},x_{2l})$.

It can be easily shown that the bound in \eqref{nuppern} is maximized when the inputs are jointly Gaussian, i.e.,  $[X_{1l},X_{2l}] \sim \mc{N}(\bf{0},\bf{K}_{P_\textit{l}})$, ${\bf{K}_{P_\textit{l}}} \in \mc{K}_{P_\textit{l}}$ with $\mc{K}_{P_\textit{l}} = \Big \{ {\bf{K}_{P_\textit{l}}} : {\bf{K}_{P_\textit{l}}}=  \left [
 \begin{smallmatrix}
  P_{1l} & \psi_\textit{l }\sqrt{P_{1l} P_{2l}}\\
\psi_\textit{l }\sqrt{P_{1l} P_{2l}} & P_{2l}
 \end{smallmatrix} \right ], -1\le \psi_\textit{l} \le {1} \Big\}$,  $l=1,\hdots,L$ with the covariance matrices $\mathbb{E}[X_{1[1,L]}X_{1[1,L]}^T]$ and $\mathbb{E}[X_{2[1,L]}X_{2[1,L]}^T]$ satisfying \eqref{p_con1} and \eqref{p_con2} respectively \cite{19},\cite{21}.
 
Next, using the specified Gaussian inputs, and proceeding as in \cite{21,khisiti}, the evaluation of the upper bound \eqref{nuppern} minimized over all possible correlations between $Y'_{l}, Y'_{2l}$, for $l=1,\hdots,L$ yields
 \begin{eqnarray}
 \label{lab1}
 R_e \le \max_{\{{\bf{K}_{P_\textit{l}}} \in {{\mc{K}}_{P_\textit{l}}}\}_{\textit{l}=1,\hdots, L}}\sum_{l=1}^L  I( X_{1l}, X_{2l};Y_l)-  I( X_{1l}, X_{2l}; Y_{2l}).
  \label{ubm}
     \end{eqnarray}
 This concludes the proof.
 \end{IEEEproof}
\vspace{0.5em}
\noindent The computation of the upper bound \eqref{guu} is given in Appendix \ref{app2}.
\vspace{0.5em}
\begin{remark}
Viewing our Gaussian model \eqref{gchan} as a specific MIMO relay-eavesdropper channel (i.e., one without interference), one can establish a genie-aided upper bound on the secrecy capacity of the model \eqref{gchan} by using recent results on MIMO wiretap channels \cite{17,19,21}, by upper bounding the secrecy rate that can be conveyed by the source and relay to the legitimate receiver on the multi-access part of the channel with that of an interference-free MIMO wiretap channel with $2L$-transmit antenna at the sender, $L$-receive antenna at the legitimate receiver and  $L$-receive antenna at the eavesdropper. However, in contrast to \eqref{guu}, the upper bound obtained this way does not show any degree of separability. More specifically, using \cite{19,21,17}, one can argue that the following is an upper-bound on the secrecy capacity of the model \eqref{gchan}, 
\begin{eqnarray}
{R}_e\le I(X_{1[1,L]},X_{2[1,L]};Y_{[1,L]}) - I(X_{1[1,L]},X_{2[1,L]};Y_{2[1,L]})								 	
\label{UpperBound__a-la-Khisti-Oggier__GaussianModel}
\end{eqnarray}
for some $ [X_{1[1,L]} X_{2[1,L]}] \sim \mc{N}(\bf{0},\bf{K}_{P})$, and  ${\bf{K}_{P}}  =  \mathbb{E}[(X_{1[1,L]}X_{2[1,L]})(X_{1[1,L]}X_{2[1,L]})^T]$ has diagonal entries that satisfies \eqref{p_con1} and \eqref{p_con2} respectively.

Because the equivalent MIMO channel is interference-free, the upper bound \eqref{UpperBound__a-la-Khisti-Oggier__GaussianModel} can be written equivalently as 
\begin{eqnarray}
R_e \leq \sum_{l=1}^L  I(X_{1l},X_{2l} ;Y_{l}|Y^{l-1}) - I(X_{1l},X_{2l} ;Y_{2l}|Y_2^{l-1}).
\label{EquivalentForm__UpperBound__a-la-Khisti-Oggier__GaussianModel}
\end{eqnarray}
Now, observe that \eqref{EquivalentForm__UpperBound__a-la-Khisti-Oggier__GaussianModel} does not show any degree of separability as in \eqref{guu}, basically because of the additional conditioning on $Y_2^{l-1}$, for $l=1,\hdots,L$. 

Also, investigating our proof in the Gaussian case, one can see that the RHS of \eqref{nupper} and its proof are fundamental. As mentioned in the proof, we could obtain the final form \eqref{ubm} essentially because the upper bound \eqref{nupper} that we established depends on the conditional joint distribution $p(y_{l},y_{2l}|x_{1l},x_{2l})$ only through its marginals.  
\end{remark}

\vspace{0.5em}

\textit{Example Application:} We consider a parallel relay channel with interference at the eavesdropper. The received signals at the relay, destination and eavesdropper are given by 
\begin{align}
\label{gchan2}
{Y_{1l,i}}&= {X_{1l,i}}+{Z_{1l,i}}\notag\\
{Y_{l,i}} &= {X_{1l,i}}+\sqrt{\rho_{1l}}{X_{2l,i}}+{Z_{l,i}}\notag\\
{Y_{2l,i}}&= {X_{1l,i}}+\sqrt{\rho_{2l}}{X_{2l,i}}+\underbrace{\sum_{k=1, k\ne l}^L {X_{1k,i}}+\sqrt{\rho_{2k}}{X_{2k,i}}}_{\text{interference}}+{Z_{2l,i}}.
\end{align}
This model can represent the equivalent channel model of a MIMO relay-eavesdropper channel with the interference at the relay and legitimate receiver avoided through singular-value decomposition; as the source can always get some feedback from both the relay and legitimate receiver, and the relay from the legitimate receiver, which then transforms the MIMO transmission into one on parallel channels among the source, relay and legitimate receiver. The eavesdropper however does not feedback information on his channel, and so is subjected to cross-antenna interference. Constraining the eavesdropper to treat the cross-antenna interference as independent noise, one can obtain an upper bound on the secrecy capacity of the model with constrained eavesdropper by direct application of \eqref{guu}. Straightforward algebra gives
\begin{align}
\label{uconn}
R_e \le& \max_{
\substack{
\sum_{l=1}^L P_{1l} \le P_1,\\ \sum_{l=1}^L P_{2l} \le P_2, \\
-1 \le {\psi_l} \le 1 \\  \text{for $l=1,\hdots,L$}}}\sum_{l=1}^L
\mathcal{C}\Big(\frac{P_{1l}+\rho_{1l}P_{2l}+2\psi_l\sqrt{\rho_{1l}P_{1l}P_{2l}}}{\sigma_{l}^2}\Big)\notag\\&- \mathcal{C}\Big(\frac{P_{1l}+\rho_{2l}P_{2l}+2\psi_l\sqrt{\rho_{2l}P_{1l}P_{2l}}}{\sum_{ k=1,k\ne l}^L P_{1k}+\sqrt{\rho_{2k}}P_{2k}+2\psi_k\sqrt{\rho_{2k}P_{1k}P_{2k}}+\sigma_{2l}^2}\Big).
\end{align}
\noindent {Then, it is clear that the upper bound \eqref{uconn} holds also for the model \eqref{gchan2} with a non-constrained eavesdropper. }

\subsection{Secrecy Capacity in Some Special Cases}
We now study the case in which the S-R links are very noisy, i.e., the relay does not hear the source.
\vspace{0.5em}
\begin{theorem}
\label{special}
For the model \eqref{gchan}, if the relay does not hear the source:
\begin{enumerate}

\item An upper bound on the perfect secrecy rate is given by 
\begin{eqnarray}
\label{equpbound}
R_e^{\textrm{up}}  =  \max \sum_{l=1}^L \mc{C} \Big(\frac{P_{1l}}{\sigma_{l}^2}\Big)- \mc{C} \Big(\frac{P_{1l}}{\sigma_{2l}^2+\rho_{2l}P_{2l}}\Big) 
\end{eqnarray}
where the maximization is over $\{P_{1l},P_{2l}\}$,   $l=1,\hdots, L$, such that $\sum_{l=1}^L P_{1l} \leq P_1$ and $\sum_{l=1}^L P_{2l} \leq P_2$.

\item A lower bound on the perfect secrecy rate is given by 
\begin{eqnarray}
\label{eqbound}
R_e^{\textrm{low}} =  \max \sum_{l=1}^L \mc{C} \Big(\frac{P_{1l}}{\sigma_{l}^2}\Big)- \mc{C} \Big(\frac{P_{1l}}{\sigma_{2l}^2+\rho_{2l}P_{2l}}\Big) 
\end{eqnarray}
where the maximization is over $\{P_{1l},P_{2l}\}$,  $l=1,\hdots, L$,   such that $\sum_{l=1}^L P_{1l} \leq P_1$, $\sum_{l=1}^L P_{2l} \leq P_2$ and
\begin{eqnarray}
\label{condition}
\sum_{l=1}^L \mathcal{C}\Big(\frac{\rho_{1l}P_{2l}}{P_{1l}+{\sigma}_{l}^2}\Big) \ge \sum_{l=1}^L  \mc{C} \Big(\frac{\rho_{2l}P_{2l}}{\sigma_{2l}^2}\Big).
\end{eqnarray}
\end{enumerate}
\end{theorem}
\vspace{0.5em}
\begin{IEEEproof}\\
\textit{\textbf{Upper Bound.}}
The bound in \eqref{equpbound} is established as follows. Our approach borrows elements from an upper bounding technique that is used in \cite{16}, and can be seen as an extension of it to the case of parallel relay-eavesdropper channels. Assume that all links between the relay and the destination are noiseless, and the eavesdropper is constrained to treat the relay's signal as unknown noise. As mentioned in \cite{16}, any upper bound for this model with full relay-destination cooperation and constrained eavesdropper also applies to the model of Theorem \ref{special}.

Now, for the model with full relay-destination cooperation and constrained eavesdropper, we develop an upper bound on the secrecy rate as follows. In this case, the destination can remove the effect of the relay transmission (which is independent from the source transmission as the relay does not hear the source), and the equivalent channel to the destination can be written as
\begin{align}
{Y'_{l,i}} &= {X_{1l,i}}+{Z_{l,i}}.
\end{align}
The eavesdropper is constrained in the sense that it is \textit{restricted} not to decode the relay's signals. Mathematically, this can be stated as follows. Let ${Z}'_{2l}$ be a random variable that has the same distribution as $X_{2l}$ and $P_{Z'_{2l}}(z) = P_{X_{2l}}(z)$, and represents unknown noise  at the eavesdropper. The channel output at the constrained
eavesdropper is given by
\begin{align}
{Y'_{2l,i}}&= {X_{1l,i}}+\underbrace{\sqrt{\rho_{2l}} {Z}'_{2l,i}}_{\text{unknown noise}}+{Z_{2l,i}}.
\end{align}
For the constrained eavesdropper the relay's transmission acts as unknown noise, with the worst case obtained with ${Z}'_{2l}$ being Gaussian, for $l=1,\hdots,L$. The rest of the proof follows by simply observing that the resulting model (with the worst case relay transmission to the eavesdropper and full relay-destination cooperation) is, in fact, a parallel Gaussian wiretap channel, the secrecy capacity of which is established in \cite{15}, i.e.,
\begin{eqnarray}
\label{newup2}
C_s =  \max \sum_{l=1}^L  I(X_{1l};Y'_{l})-I(X_{1l};Y'_{2l})
\end{eqnarray}
where the maximization is over $X_{1l}\sim \mc{N}(0,P_{1l})$ and $X_{2l} \sim \mc{N}(0,P_{2l})$, $l=1,\hdots, L$, with $\sum_{l=1}^{L}P_{1l} \leq P_1$ and $\sum_{l=1}^{L}P_{2l} \leq P_2$.

Finally, straightforward algebra which is omitted for brevity shows that the computation of \eqref{newup2} gives \eqref{equpbound}.

\vspace{0.5em}
\textit{\textbf{Lower Bound.}}
The proof of the lower bound follows by evaluating the equivocation in Theorem~\ref{low} with a specific choice of the variables. More specifically, evaluating \eqref{innerd} with the choice $|\mc{A}^c|:=L$, $V_{1l}:=X_{1l}$, $V_{2l}:=X_{2l}$,  with $X_{1l}\sim\mc N(0,P_{1l})$ independent of $X_{2l}\sim\mc N(0,P_{2l})$, $l=1,\hdots,L$ and such that \eqref{condition} is satisfied, we get the rate expression in the RHS of \eqref{eqbound}. The RHS of \eqref{eqbound} then follows by maximization over all $\{P_{1l},P_{2l}\}$, $l=1,\hdots,L$, satisfying \eqref{condition} and the total power constraints $\sum_{l=1}^{L}P_{1l} \leq P_1$ and $\sum_{l=1}^{L}P_{2l} \leq P_2$. 
\end{IEEEproof}

\vspace{0.5em}
\begin{remark}
The upper \eqref{equpbound} and lower \eqref{eqbound} bounds on the perfect secrecy rate of Theorem 4 have same expressions but are maximized over different input sets. These bounds coincide  \textit{only} when the inputs ($\{P_{1l},P_{2l}\}$) that maximize  the upper bound \eqref{equpbound} also satisfy \eqref{condition}. For this specific case, perfect secrecy is established and is given  by 
\begin{eqnarray}
\label{capacity}
 C_s = \max \sum_{l=1}^L \mc{C} \Big(\frac{P_{1l}}{\sigma_{l}^2}\Big)- \mc{C} \Big(\frac{P_{1l}}{\sigma_{2l}^2+\rho_{2l}P_{2l}}\Big) 
\end{eqnarray}
where the maximization is over $\{P_{1l},P_{2l}\}$,  $l=1,\hdots, L$, such that $\sum_{l=1}^L P_{1l} \leq P_1$,  $\sum_{l=1}^L P_{2l} \leq P_2$ and 
\begin{eqnarray}
\sum_{l=1}^L \mathcal{C}\Big(\frac{\rho_{1l}P_{2l}}{P_{1l}+{\sigma}_{l}^2}\Big) \ge \sum_{l=1}^L  \mc{C} \Big(\frac{\rho_{2l}P_{2l}}{\sigma_{2l}^2}\Big).
\end{eqnarray}
\end{remark}
\vspace{0.5em}

\section{Example Application}
In this section we apply the results which we established for the Gaussian memoryless model in section III to study a fading relay-eavesdropper channel. 

For a fading relay-eavesdropper channel, the received signals at the relay, legitimate receiver and eavesdropper are given by
\begin{align}
\label{raych}
{Y_{1,i}}&= {h_{sr,i}{X_{1,i}}}+{Z_{1,i}}\notag\\
{Y_{i}}  &= {h_{sd,i}X_{1,i}}+{h_{rd,i}X_{2,i}}+{Z_{i}}\notag\\
{Y_{2,i}}&= {h_{se,i}X_{1,i}}+{h_{re,i}X_{2,i}}+{Z_{2,i}}
\end{align}
where $i$ is the time index, $ h_{sd,i}$, $h_{rd,i}$, $h_{se,i}$, $h_{re,i}$ and  $h_{sr,i}$ are the fading gain coefficients associated with S-D, R-D, S-E, R-E and S-R links, given by complex Gaussian random variables with zero mean and unit variance respectively. The noise processes $\{Z_{1,i}\},\{Z_{i}\},\{Z_{2,i}\}$ are zero mean i.i.d complex Gaussian random variables with  variances $\sigma_{1}^2$, $\sigma^2$ and $\sigma_{2}^2$ respectively. The source and relay input sequences are subject to an average power constraint, i.e., $\sum_{i=1}^n \mathbb{E}[\parallel X_{1,i}\parallel ^2] \le nP_1$, $\sum_{i=1}^n \mathbb{E}[\parallel X_{2,i}\parallel^2] \le nP_2$. We define $\bar{h}_i := [h_{sd,i} \hspace{.5em} h_{rd,i} \hspace{.5em}  h_{se,i} \hspace{.5em}   h_{re,i} \hspace{.5em}  h_{sr,i}]$ and assume that  perfect non-causal channel state information (CSI) is available at all nodes. For a given fading state realization $\bar{h}_i$, the fading relay-eavesdropper channel is a Gaussian relay-eavesdropper channel. Therefore, for a given channel state with $L$ fading state realizations, i.e., $\bar{h}=\{\bar{h}_i\}_{i=1}^L$, the fading relay-eavesdropper channel can be seen as a parallel Gaussian relay-eavesdropper channel with $L$ subchannels. The power allocation vectors at the source and relay are denoted by $P_1(\bar{h})$ and $P_2(\bar{h})$ respectively. The ergodic achievable secrecy rate of the fading relay-eavesdropper channel \eqref{raych}, which follows from \eqref{glowm} is given by 
 { \small{
\begin{align}
\label{fcap}
{R}_{e}^{\textrm{low}} =&
\max_{
\substack{
\mathbb{E}[P_1(\bar{h})] \le P_1,\\ \mathbb{E}[P_2(\bar{h})] \le P_2,\\
0 \le\alpha(\bar{h})\le 1}} \min \Big \{ \mathbb{E}_{\bar{h}\in \mc{A}} \Big[2\mathcal{C}\Big(
\frac{|{h}_{sd}|^2P_1(\bar{h})+ |{h}_{rd}|^2P_2(\bar{h})+2\sqrt{\bar{\alpha}(\bar{h})|{h}_{sd}|^2P_1(\bar{h})|h_{rd}|^2P_2(\bar{h})}}{{\sigma}^2}\Big)\notag\\&-2\mathcal{C}\Big(\frac{|{h}_{se}|^2P_1(\bar{h})+|{h}_{re}|^2P_2(\bar{h})+2\sqrt{\bar{\alpha}(\bar{h})|{h}_{se}|^2P_1(\bar{h})|{h}_{re}|^2P_2(\bar{h})}}{{\sigma}_{2}^2}\Big)\Big]^+, \mathbb{E}_{\bar{h} \in \mc{A}} \Big[2\mathcal{C} \Big( \frac{\alpha(\bar{h})|{h}_{sr}|^2P_1(\bar{h})}{\sigma_{1}^2}\Big)\notag\\&-2\mathcal{C}\Big(\frac{|{h}_{se}|^2P_1(\bar{h})+|{h}_{re}|^2P_2(\bar{h})+2\sqrt{\bar{\alpha}(\bar{h})|{h}_{se}|^2P_1(\bar{h})|{h}_{re}|^2P_2(\bar{h})}}{{\sigma}_{2}^2}\Big)\Big]^+\Big\} \notag\\&+ \min\Big\{ \mathbb{E}_{\bar{h} \in \mc{A}^c} \Big [2\mathcal{C} \Big( \frac{|{h}_{sd}|^2{P_1(\bar{h})}+|{h}_{rd}|^2{P_2(\bar{h})}}{\sigma^2}\Big)-2\mathcal{C} \Big( \frac{|{h}_{se}|^2{P_1(\bar{h})}+|{h}_{re}|^2{P_2(\bar{h})}}{\sigma_{2}^2}\Big)\Big]^+,\notag\\&\mathbb{E}_{\bar{h} \in \mc{A}^c} \Big[ 2\mathcal{C}\Big(\frac{|{h}_{sd}|^2P_1(\bar{h})}{\sigma^2}\Big) + 2\mathcal{C}\Big(\frac{|{h}_{re}|^2P_2(\bar{h})}{\sigma_{2}^2}\Big)-  2\mathcal{C} \Big( \frac{|{h}_{se}|^2{P_1(\bar{h})}+|{h}_{re}|^2{P_2(\bar{h})}}{{\sigma}_{2}^2}\Big)\Big ]^+\Big \}.
\end{align}}}
\normalsize
\noindent {The upper bound for the fading relay-eavesdropper channel \eqref{raych} follows directly from the upper bound established for the parallel Gaussian relay-eavesdropper channel \eqref{guu}. Straightforward algebra which is omitted for brevity gives}

{\small {
\begin{align}
\label{eqfg}
R_e^{\textrm{up}} =\max_{
\substack{
\mathbb{E}[P_1(\bar{h})] \le P_1,\\ \mathbb{E}[P_2(\bar{h})] \le P_2,\\
-1 \le \psi(\bar{h}) \le 1}} &\mathbb{E}_{\bar{h}} \Big \{
2\mc{C}\Big(\frac{|h_{sd}|^2 P_{1}(\bar{h})+|h_{rd}|^2P_{2}(\bar{h})+2\psi(\bar{h}) \sqrt{|h_{sd}|^2P_{1}(\bar{h})|h_{rd}|^2P_{2}(\bar{h})}}{\sigma^2}\Big)\notag\\&- 2\mc{C}\Big(\frac{|h_{se}|^2 P_{1}(\bar{h})+|h_{re}|^2P_{2}(\bar{h})+2\psi(\bar{h}) \sqrt{|h_{se}|^2P_{1}(\bar{h})|h_{re}|^2P_{2}(\bar{h})}}{\sigma_{2}^2}\Big) \Big\}.
\end{align}
 }}


\section{Numerical Results}
In this section we provide numerical examples to illustrate the performance of fading relay-eavesdropper channel. We consider a fading relay-eavesdropper channel with \textit{L} realizations of fading state. It is assumed that perfect channel state information is available at all nodes. We can consider this channel as a Gaussian relay-eavesdropper channel with \textit{L} subchannels. Alternatively, this model can be seen as an OFDM system with $L$ sub-carriers. We model channel gain between node $i \in \{s,r\}$ and $j \in \{r,d,e\}$ as distance dependent Rayleigh fading, that is, $h_{i,j} = h'_{i,j}{d_{i,j}^{-\gamma/2}}$,
where $\gamma$ is the path loss exponent, $d_{i,j}$ is the distance between the node $i$ and $j$, and $h'_{i,j}$ is a complex Gaussian random variable with zero mean and variance one. Each subchannel is corrupted by additive white Gaussian noise with zero mean and variance one. Furthermore, for each symbol transmission same subchannel is used on S-R and R-D links to make the optimization tractable. The objective function for both lower and upper bounds  are optimized numerically using AMPL with a commercially available solver, for instance SNOPT.

\begin{figure}[h]
\centering
\includegraphics[width=\linewidth]{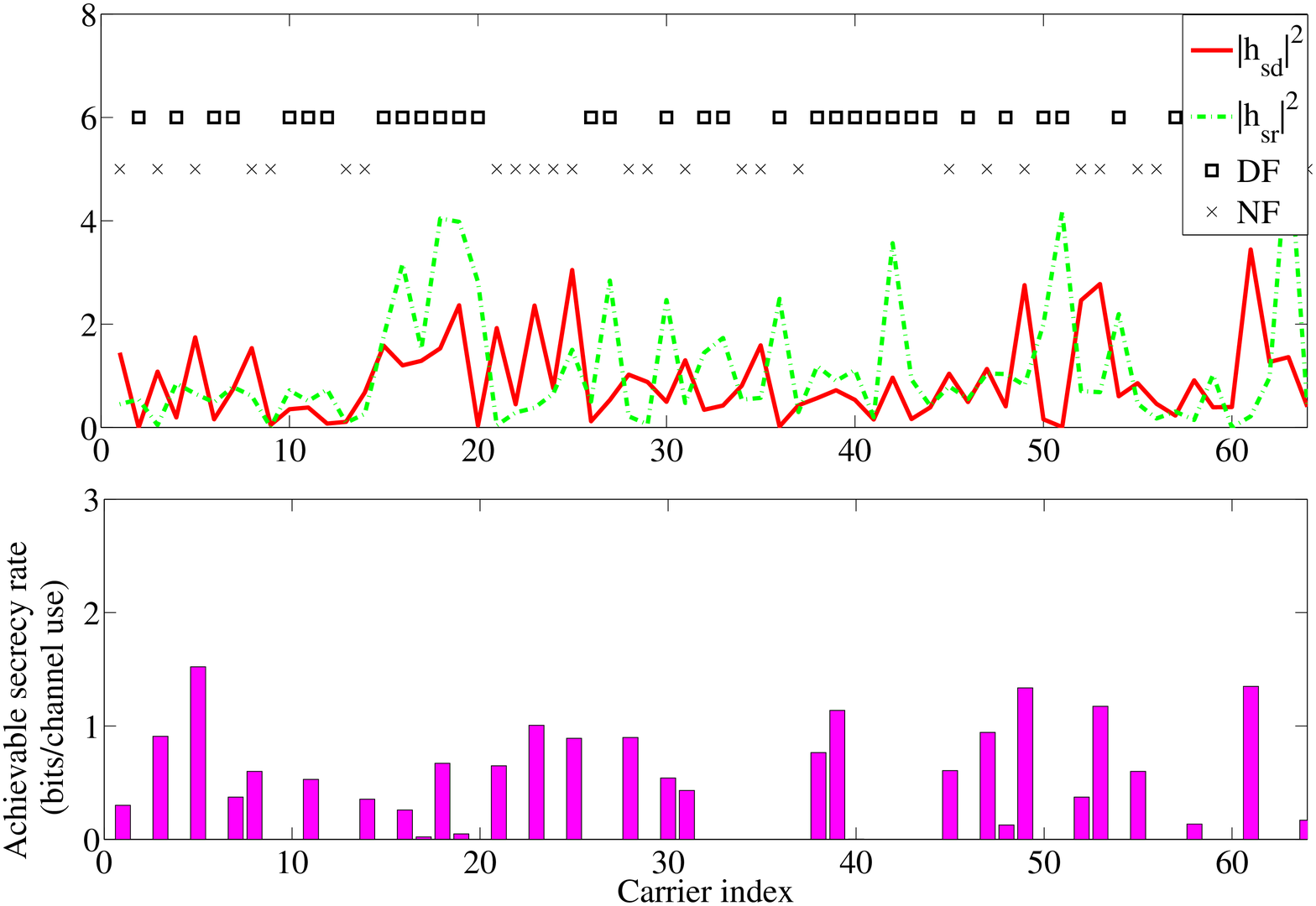}
 \caption{Achievable perfect secrecy rate of a parallel relay-eavesdropper channel.}\label{pcp}
\end{figure}

To illustrate the system performance, we set the source and relay power to 64 Watt each. We consider a network geometry in which the source is located at the point (0,0), the relay is located at the point ($d$,0), the destination is located at the point (1,0) and the eavesdropper is located at the point (0,1), where $d$ is the distance between the source and the relay. In all numerical results we set path loss exponent $\gamma$:=2 and $L:=64$. For all numerical examples, secrecy rate is given by bits per channel use. For each subchannel the selection of the coding scheme at the relay is based on the relative strength of the S-D link w.r.t the S-R link, i.e., we use NF scheme (set $\mc{A}^c$) when $|h_{sd}|^2\ge|h_{sr}|^2$  and DF scheme (set $\mc{A}$) when $|h_{sd}|^2<|h_{sr}|^2$. Fig. \ref{pcp} shows the power allocation for a fading channel with 64 subchannels  where the relay is located at (0.5,0), and  marker `$\times$' denotes NF on a particular subchannel while marker `$\square$' denotes DF on a particular subchannel. It can be seen from Fig. \ref{pcp} that, achievable perfect secrecy rate is zero for some subchannels. Roughly speaking, this happens when the condition $|h_{rd}|^2>|h_{re}|^2$  is violated.

\begin{figure}[ht]
\centering
\includegraphics[width=\linewidth]{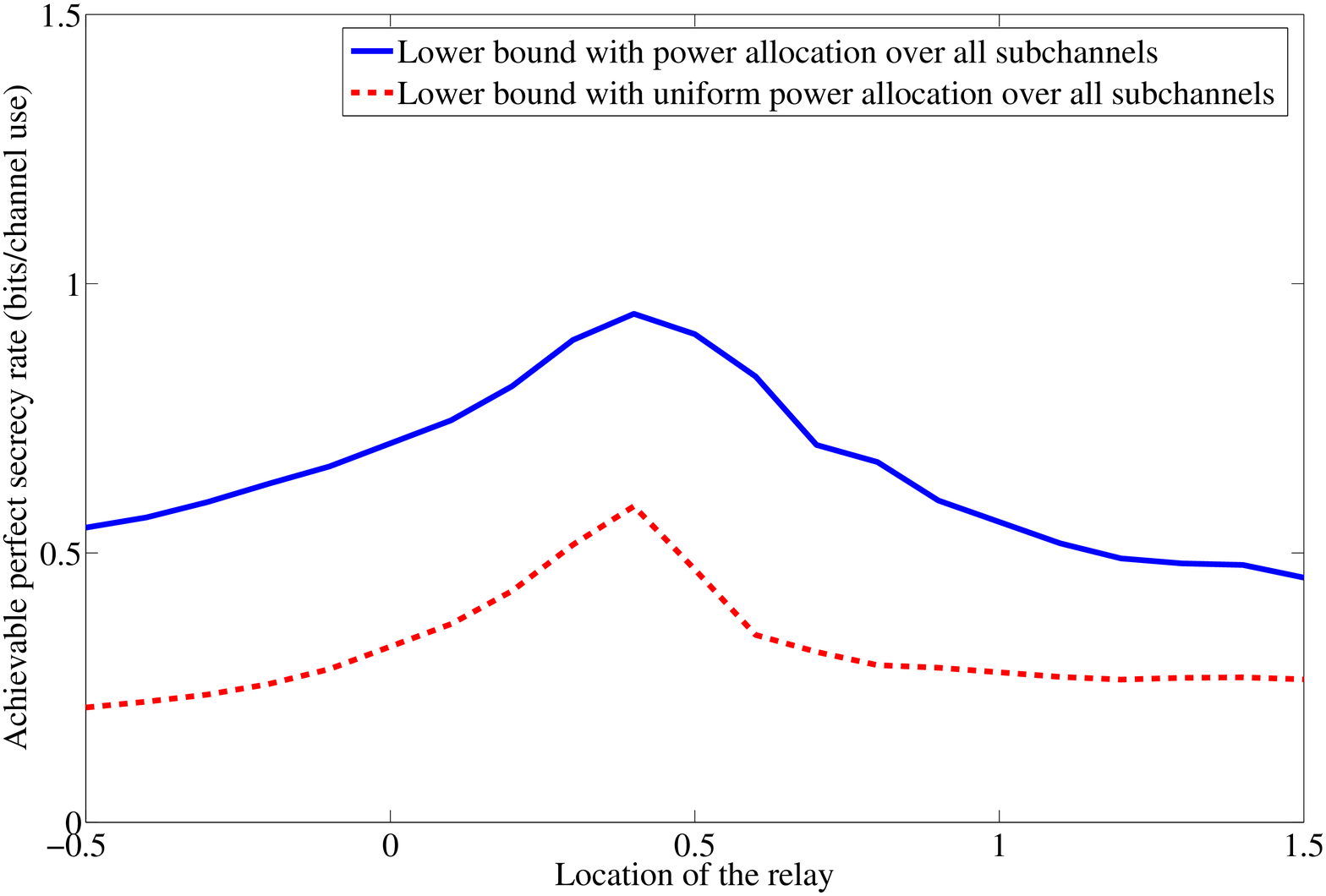}
\caption{Comparison of achievable perfect secrecy rate of the lower bound with optimized power allocation and with uniform power allocation over all subchannels.}
 \label{fig4}
 \end{figure}

Fig. \ref{fig4} compares the average perfect secrecy rate of the lower bound, with optimized power allocation  and with uniform power allocation, i.e., allocating same power at the source and relay for all subchannels in $\bar{h} \in \mc{A}$ and in $\bar{h} \in \mc{A}^c$. It can be seen that for separate source and relay powers, optimized power allocation scheme outperforms uniform power allocation scheme. This fact follows because optimized power allocation scheme maximizes the achievable perfect secrecy rate and hence enhances the system performance.
 \begin{figure}[ht]
\centering
\includegraphics[width=\linewidth]{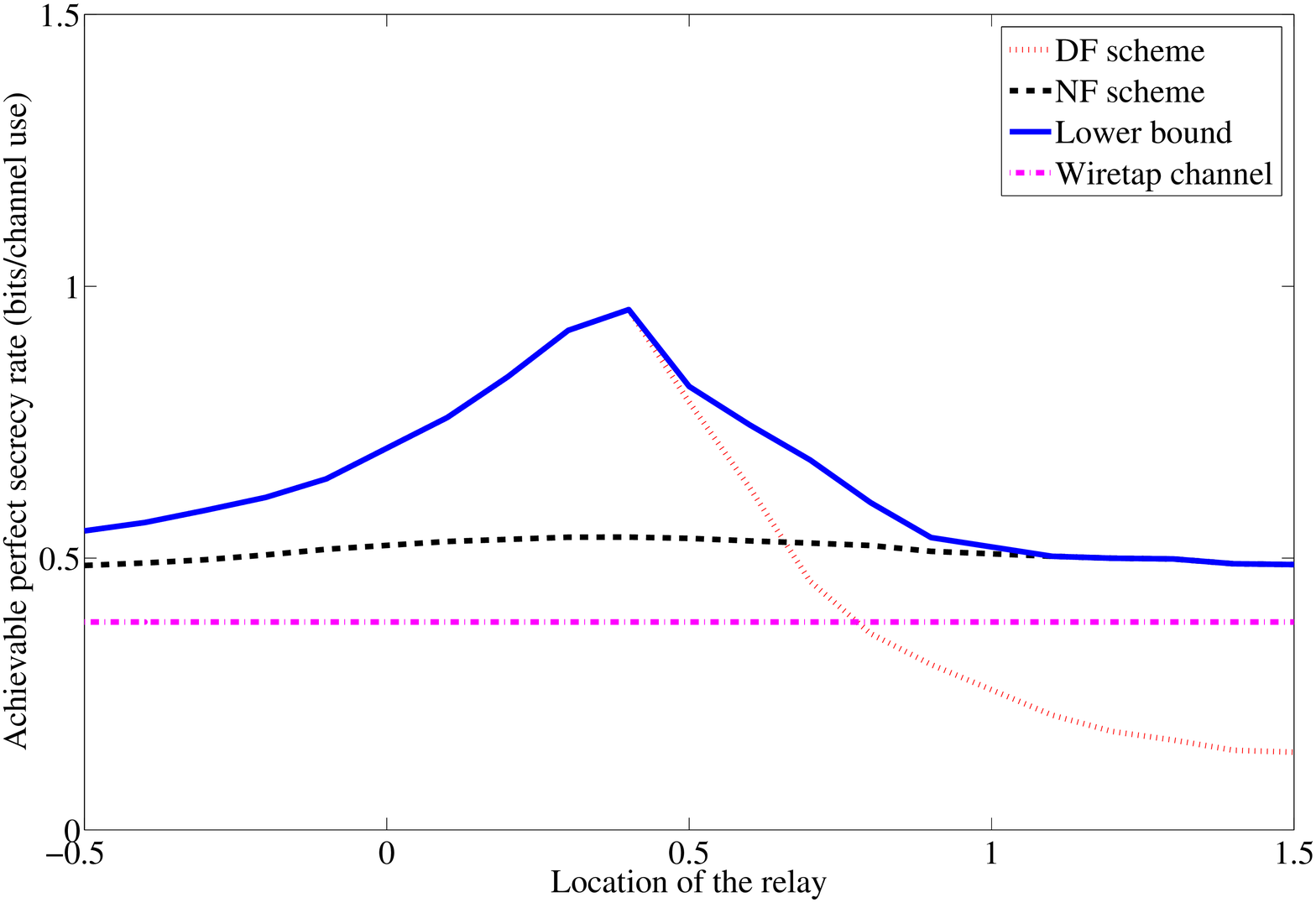}
\caption{Comparison of achievable perfect secrecy rate of some schemes with the lower bound.}
 \label{fig5}
 \end{figure}

Mode selection at the relay by only considering the relative strength of the S-D and the S-R link in the lower bound is suboptimal because the achievable secrecy rate \eqref{fcap} also depends on the gain of other link. We now consider the case in which the relay selects the scheme which maximizes the rate for each subchannel. We plot the lower bound with this criteria and compare it with the case in which same scheme is used on all subchannels. As a reference we consider the case in which there is no relay, i.e., a parallel  wiretap channel. Fig. \ref{fig5} shows the achievable average perfect secrecy rate of different schemes. It can be seen that when the relay is close to the source, DF scheme on all subchannels gives higher secrecy rate. Similarly when the relay is close to the destination, NF scheme on all subchannels offers better rate. The region when the relay is between $0.5 < d < 1.2$ is of particular interest. In this region the relay selects between DF scheme and NF scheme for each subchannel and utilizes the gain from both schemes. It is interesting to note that when the relay is close to the destination, use of DF scheme on all subchannels does not offer any gain because in this case the relay is unable to decode the source codewords and hence the average secrecy rate decreases. The lower bound always perform better than the wiretap channel which shows the usefulness of the relay.

 \begin{figure}[ht]
\centering
\includegraphics[width=\linewidth]{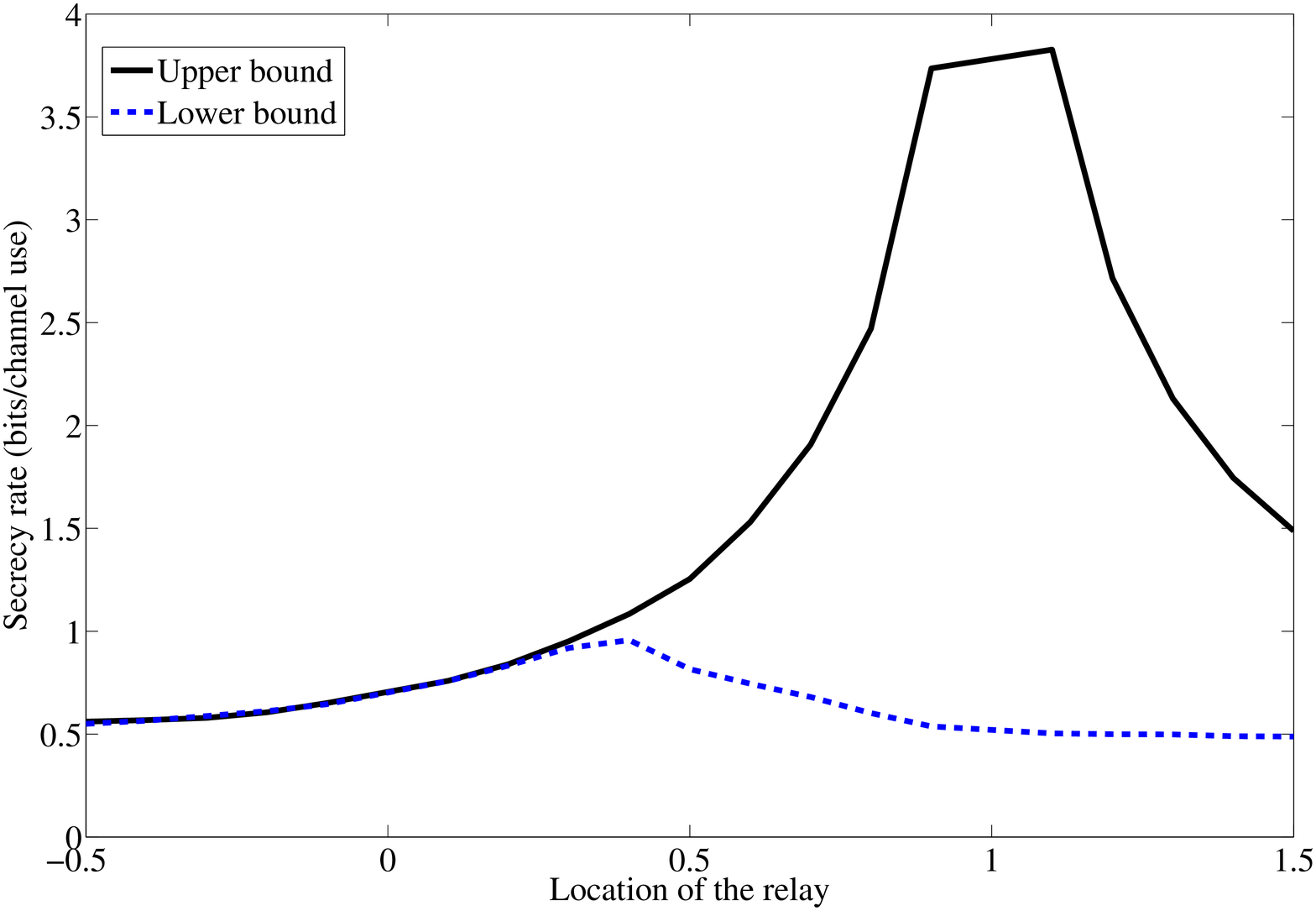}
\caption{Bounds on perfect secrecy rate.}
 \label{fig6}
 \end{figure}
In Fig. \ref{fig6} we compare the lower bound obtained in Fig. \ref{fig5}, with the upper bound on the secrecy capacity for the fading relay-eavesdropper channel. It can be seen that when the relay is close to the source, the lower and upper bounds coincide. This is achieved by using DF scheme on all subchannels.

\section{Conclusions}
We studied the problem of secure communication over parallel relay channel. Outer and inner bounds on the rate-equivocation are established for the DM case. Developing an outer bound on the parallel relay-eavesdropper channel is non-trivial and it does not follow directly from the one established in \cite{10}. For the Gaussian memoryless case, lower and upper bounds on the perfect secrecy rate are established. The computable upper bound for the Gaussian model shows some separability over subchannels. In the case in which the relay does not hear the source, under some specific conditions the lower and upper bounds coincide and secrecy capacity is established. We apply the results established for the Gaussian memoryless model to a more practical fading relay-eavesdropper channel. Numerical examples showed that power adjustment among parallel channels results in higher secrecy rate.

\appendices
\section{{Proof of Theorem 1}}
\label{app1}
The proof generalizes the results of Theorem 1 in \cite{10} and uses elements from a similar proof in the context of parallel BCC in \cite{15}.\\
\textbf{a)} We first bound the equivocation rate as follows.
\setlength{\arraycolsep}{0.2em}
\begin{eqnarray}
\label{upper}
nR_{e}&=& {H}(W \mid Y_{2[1,L]}^n)\notag \\
&=& H({W})- I(W;Y_{2[1,L]}^n)\notag\\
&=& I(W;Y_{[1,L]}^n)-I(W;Y_{2[1,L]}^n)+{H}(W \mid Y_{[1,L]}^n)\notag \\
&\overset{(a)}{\le}& I(W;Y_{[1,L]}^n)-I(W;Y_{2[1,L]}^n)+n\epsilon_{n} \notag\\
&=& \sum_{l=1}^L I(W;Y_l^n\mid Y_{[1,l-1]}^n)-I(W;Y_{2l}^n\mid Y_{2[l+1,L]}^n)+n\epsilon_{n} \notag\\
&{=}& \sum_{l=1}^L \sum_{i=1}^n I(W;Y_{li}\mid Y_l^{i-1},Y_{[1,l-1]}^n)- I(W;Y_{2li}\mid Y_{2l[i+1]}^n, Y_{2[l+1,L]}^n)+n\epsilon_{n}\notag\\
&{=}& \sum_{l=1}^L \sum_{i=1}^n I(W, Y_{2l[i+1]}^n, Y_{2[l+1,L]}^n;Y_{li}\mid Y_l^{i-1},Y_{[1,l-1]}^n)- I(Y_{2l[i+1]}^n, Y_{2[l+1,L]}^n;Y_{li}\mid W, Y_l^{i-1},Y_{[1,l-1]}^n)\notag\\&&- I(W, Y_l^{i-1},Y_{[1,l-1]}^n;Y_{2li}\mid Y_{2l[i+1]}^n, Y_{2[l+1,L]}^n)+ I(Y_l^{i-1},Y_{[1,l-1]}^n;Y_{2li}\mid W, Y_{2l[i+1]}^n, Y_{2[l+1,L]}^n)+n\epsilon_{n}\notag \\
&\overset{(b)}{=}& \sum_{l=1}^L \sum_{i=1}^n I(W, Y_{2l[i+1]}^n, Y_{2[l+1,L]}^n;Y_{li}\mid Y_l^{i-1},Y_{[1,l-1]}^n)- I(W, Y_l^{i-1},Y_{[1,l-1]}^n;Y_{2li}\mid Y_{2l[i+1]}^n, Y_{2[l+1,L]}^n)\notag\\&&+n\epsilon_{n}\notag \\
&=& \sum_{l=1}^L \sum_{i=1}^n I(Y_{2l[i+1]}^n, Y_{2[l+1,L]}^n;Y_{li}\mid Y_l^{i-1},Y_{[1,l-1]}^n) +I(W ;Y_{li}\mid  Y_l^{i-1},Y_{[1,l-1]}^n, Y_{2l[i+1]}^n, Y_{2[l+1,L]}^n)\notag\\&&- I(Y_l^{i-1},Y_{[1,l-1]}^n;Y_{2li}\mid Y_{2l[i+1]}^n, Y_{2[l+1,L]}^n)-I(W;Y_{2li}\mid Y_l^{i-1},Y_{[1,l-1]}^n, Y_{2l[i+1]}^n, Y_{2[l+1,L]}^n)+n\epsilon_{n}\notag \\
&\overset{(c)}{=}& \sum_{l=1}^L \sum_{i=1}^n I(W ;Y_{li}\mid  Y_l^{i-1},Y_{[1,l-1]}^n, Y_{2l[i+1]}^n, Y_{2[l+1,L]}^n)-I(W;Y_{2li}\mid Y_l^{i-1},Y_{[1,l-1]}^n, Y_{2l[i+1]}^n, Y_{2[l+1,L]}^n)\notag\\&&+n\epsilon_{n}
\end{eqnarray}
where $\epsilon_n \rightarrow 0$ as $n \rightarrow \infty$;
$(a)$ follows from Fano's inequality; and
$(b)$ and $(c)$ follows from lemma 7 in \cite{9}.
 
We introduce a random variable $T$ uniformly distributed over $\{1,2,\hdots,n\}$ and set, $U_{li} :=  Y_l^{i-1},Y_{[1,l-1]}^n, Y_{2l[i+1]}^n, Y_{2[l+1,L]}^n$, $V_{1li} :=  W, Y_{2l[i+1]}^n, Y_{2[l+1,L]}^n$ and $V_{2li} :=  Y_l^{i-1},Y_{[1,l-1]}^n$. We define  $U_{l} = (T,U_{li}), V_{1l} = (T,V_{1li}), V_{2l} = (T,V_{2li}), X_{1l}=X_{1T}, X_{2l}=X_{2T},Y_{l}=Y_{T}, Y_{1l}=Y_{1T}, Y_{2l}=Y_{2T}$, for $l=1,\hdots,L$.
Note that $(U_l,V_{1l},V_{2l},X_{1l},X_{2l},Y_{l},Y_{1l},Y_{2l})$ satisfies the following Markov chain condition
\begin{equation}
{U_l}\rightarrow({V_{1l},V_{2l}})\rightarrow({X_{1l},X_{2l}})\rightarrow({Y_l,Y_{1l},Y_{2l}}),  \text{for $l =1,\hdots,L$}.\notag
\end{equation}
Thus, we have
\begin{eqnarray}
R_{e}&\le& \frac{1}{n}\sum_{l=1}^L \sum_{i=1}^n I(W ;Y_{li}\mid  Y_l^{i-1},Y_{[1,l-1]}^n, Y_{2l[i+1]}^n, Y_{2[l+1,L]}^n) \notag\\&&-I(W;Y_{2li}\mid Y_l^{i-1},Y_{[1,l-1]}^n, Y_{2l[i+1]}^n, Y_{2[l+1,L]}^n)+\epsilon_{n}\notag \\
&=& \frac{1}{n}\sum_{l=1}^L \sum_{i=1}^n I(W, Y_l^{i-1},Y_{[1,l-1]}^n, Y_{2l[i+1]}^n, Y_{2[l+1,L]}^n ;Y_{li}\mid  Y_l^{i-1},Y_{[1,l-1]}^n, Y_{2l[i+1]}^n, Y_{2[l+1,L]}^n) \notag \\&&-I(W,Y_l^{i-1},Y_{[1,l-1]}^n, Y_{2l[i+1]}^n, Y_{2[l+1,L]}^n;Y_{2li}\mid Y_l^{i-1},Y_{[1,l-1]}^n, Y_{2l[i+1]}^n, Y_{2[l+1,L]}^n)+\epsilon_{n}\notag \\
\label{u1u}
&\overset{(d)}{=}&  \frac{1}{n} \sum_{l=1}^L \sum_{i=1}^n I(V_{1li},V_{2li} ;Y_{li}\mid U_{li})-I(V_{1li},V_{2li} ;Y_{2li}\mid U_{li})+\epsilon_{n}\\
\label{u1}
&\overset{(e)}{=}& \sum_{l=1}^L I(V_{1l},V_{2l} ;Y_{l}\mid U_{l})-I(V_{1l},V_{2l} ;Y_{2l}\mid U_{l})+\epsilon_{n}
\end{eqnarray}
where $(d)$ and $(e)$ follow by using the above definition.\\
We can also bound the equivocation rate as follows.
We continue from \eqref{upper} to get
\begin{eqnarray}
R_{e}&\le& \frac{1}{n}\sum_{l=1}^L \sum_{i=1}^n I(W ;Y_{li}\mid  Y_l^{i-1},Y_{[1,l-1]}^n, Y_{2l[i+1]}^n, Y_{2[l+1,L]}^n)\notag\\&& -I(W;Y_{2li}\mid Y_l^{i-1},Y_{[1,l-1]}^n, Y_{2l[i+1]}^n, Y_{2[l+1,L]}^n)+\epsilon_{n}\notag \\
&=& \frac{1}{n}\sum_{l=1}^L \sum_{i=1}^n I(W, Y_{2l[i+1]}^n, Y_{2[l+1,L]}^n ;Y_{li}\mid  Y_l^{i-1},Y_{[1,l-1]}^n, Y_{2l[i+1]}^n, Y_{2[l+1,L]}^n) \notag \\&&-I(W,Y_l^{i-1},Y_{[1,l-1]}^n, Y_{2l[i+1]}^n, Y_{2[l+1,L]}^n;Y_{2li}\mid Y_l^{i-1},Y_{[1,l-1]}^n, Y_{2l[i+1]}^n, Y_{2[l+1,L]}^n)+\epsilon_{n}\notag \\
&\le& \frac{1}{n} \sum_{l=1}^L \sum_{i=1}^n I(W, Y_{2l[i+1]}^n, Y_{2[l+1,L]}^n ;Y_{li},Y_{1li}\mid  Y_l^{i-1},Y_{[1,l-1]}^n, Y_{2l[i+1]}^n, Y_{2[l+1,L]}^n) \notag \\&&-I(W,Y_l^{i-1},Y_{[1,l-1]}^n, Y_{2l[i+1]}^n, Y_{2[l+1,L]}^n;Y_{2li}\mid Y_l^{i-1},Y_{[1,l-1]}^n, Y_{2l[i+1]}^n, Y_{2[l+1,L]}^n)+\epsilon_{n}\notag \\
\label{u21}
&\overset{(f)}{=}&  \frac{1}{n} \sum_{l=1}^L \sum_{i=1}^n I(V_{1li};Y_{li},Y_{1li}\mid V_{2li},U_{li})-I(V_{1li}, V_{2li} ;Y_{2li}\mid U_{li})+\epsilon_{n} \\
\label{u2}
&\overset{(g)}{=}& \sum_{l=1}^L I(V_{1l};Y_l, Y_{1l}\mid  V_{2l},U_{l})-I(V_{1l},V_{2l} ;Y_{2l}\mid U_{l})+\epsilon_{n}
\end{eqnarray}
where $(f)$ and $(g)$ follow from the above definition.

\textbf{b)} We now bound the rate $R$ as follows.
\begin{eqnarray}
\label{achrate}
nR&=& {H}(W)\notag \\
&=& I(W;Y_{[1,L]}^n)+H(W \mid Y_{[1,L]}^n)\notag\\
&\overset{(h)}{\le}&  I(W;Y_{[1,L]}^n)+ n \epsilon_{n}\notag \\
&=& \sum_{l=1}^L I(W;Y_{l}^n \mid Y_{[1,l-1]}^n)+n \epsilon_{n}\notag\\
&=&  \sum_{l=1}^L\sum_{i=1}^n I(W;Y_{li}\mid Y_l^{i-1}, Y_{[1,l-1]}^n)+ n \epsilon_{n} \notag\\
&=&  \sum_{l=1}^L\sum_{i=1}^n H(Y_{li}\mid Y_l^{i-1}, Y_{[1,l-1]}^n)-H(Y_{li}\mid W, Y_l^{i-1}, Y_{[1,l-1]}^n)+ n \epsilon_{n} \notag\\
&\overset{(i)}{\le}&  \sum_{l=1}^L\sum_{i=1}^n H(Y_{li})-H(Y_{li}\mid W, Y_l^{i-1}, Y_{[1,l-1]}^n)+ n \epsilon_{n} \notag\\
&\overset{(j)}{\le}&  \sum_{l=1}^L\sum_{i=1}^n H(Y_{li})-H(Y_{li}\mid W, Y_l^{i-1}, Y_{[1,l-1]}^n, Y_{2l[i+1]}^n, Y_{2[l+1,L]}^n)+ n \epsilon_{n} \notag\\
&=&  \sum_{l=1}^L\sum_{i=1}^n I(W, Y_l^{i-1}, Y_{[1,l-1]}^n, Y_{2l[i+1]}^n, Y_{2[l+1,L]}^n;Y_{li}) + n \epsilon_{n} \notag\\
&=&  \sum_{l=1}^L\sum_{i=1}^n I(V_{1li},V_{2li} ;Y_{li})+ n \epsilon_{n}.
\end{eqnarray}
Hence, we have
\begin{eqnarray}
\label{u3}
R&\le& \frac{1}{n} \sum_{l=1}^L\sum_{i=1}^n I(V_{1li},V_{2li} ;Y_{li})+ \epsilon_{n} \notag\\
 &\le& \sum_{l=1}^L I(V_{1l},V_{2l} ;Y_{l})+  \epsilon_{n}
\end{eqnarray}
where $(h)$ follows from Fano's inequality; $(i)$ and $(j)$ follows from the fact that conditioning reduces entropy.
We can also bound the rate $R$ as follows
\begin{eqnarray}
\label{achrate2}
nR&=& {H}(W)\notag \\
&=& I(W;Y_{[1,L]}^n)+H(W \mid Y_{[1,L]}^n)\notag\\
&\overset{(k)}{\le}&  I(W;Y_{[1,L]}^n)+ n \epsilon_{n}\notag \\
&=& \sum_{l=1}^L I(W;Y_l^n\mid Y_{[1,l-1]}^n)+ n \epsilon_{n}\notag\\
&=&    \sum_{l=1}^L\sum_{i=1}^n I(W;Y_{li}\mid Y_l^{i-1}, Y_{[1,l-1]}^n)+ n \epsilon_{n} \notag\\
&\le&  \sum_{l=1}^L\sum_{i=1}^n I(W;Y_{1li},Y_{li}\mid Y_l^{i-1}, Y_{[1,l-1]}^n)+ n \epsilon_{n} \notag\\
&=&  \sum_{l=1}^L\sum_{i=1}^n H(Y_{1li},Y_{li}\mid Y_l^{i-1}, Y_{[1,l-1]}^n)-H(Y_{1li},Y_{li}\mid W, Y_l^{i-1}, Y_{[1,l-1]}^n)+ n \epsilon_{n} \notag\\
&\overset{(l)}{\le}&  \sum_{l=1}^L\sum_{i=1}^n H(Y_{1li},Y_{li}\mid Y_l^{i-1}, Y_{[1,l-1]}^n)-H(Y_{1li},Y_{li}\mid W, Y_l^{i-1}, Y_{[1,l-1]}^n, Y_{2l[i+1]}^n, Y_{2[l+1,L]}^n)+ n \epsilon_{n} \notag\\
&=&  \sum_{l=1}^L\sum_{i=1}^n I(W, Y_{2l[i+1]}^n, Y_{2[l+1,L]}^n ;Y_{1li},Y_{li} \mid Y_l^{i-1}, Y_{[1,l-1]}^n ) + n \epsilon_{n} \notag\\
&=&  \sum_{l=1}^L\sum_{i=1}^n I(V_{1li};Y_{1li},Y_{li}\mid V_{2li})+ n \epsilon_{n}.
\end{eqnarray}
Hence, we have
\begin{eqnarray}
\label{u4}
R &\le& \frac{1}{n}\sum_{l=1}^L\sum_{i=1}^n I(V_{1li};Y_{1li},Y_{li}\mid V_{2li}) +  \epsilon_{n} \notag\\
  &\le& \sum_{l=1}^L I(V_{1l} ;Y_{l},Y_{1l}\mid V_{2l})+  \epsilon_{n}
\end{eqnarray}
where $(k)$ follows from Fano's inequality; and $(l)$ follows from the fact that conditioning reduces the entropy.
 
Therefore an outer bound on the achievable rate equivocation region is given by the following set:
\begin{eqnarray}
\bigcup \Big\{(R,R_e)\:  \text{that satisfy}\:  \eqref{u1},\eqref{u2},\eqref{u3},\eqref{u4}\Big\}
 \end{eqnarray}
where the union is over all probability distributions  $p(u_{[1,L]},v_{1[1,L]},v_{2[1,L]},x_{1[1,L]},x_{2[1,L]},y_{[1,L]},y_{1[1,L]},$\\$y_{2[1,L]})$. Finally we note that the terms in  \eqref{u1},\eqref{u2},\eqref{u3}, and \eqref{u4} depend on the probability distribution $p(u_{[1,L]},v_{1[1,L]},v_{2[1,L]},x_{1[1,L]},x_{2[1,L]},y_{[1,L]},y_{1[1,L]},y_{2[1,L]})$ only through $p(u_l,v_{1l},v_{2l},x_{1l},x_{2l},y_{l},y_{1l},y_{2l})$. Hence, there is no loss of optimality to consider only those distributions that have the form
\begin{eqnarray}
\label{mark}
\prod_{l=1}^L p(u_l,v_{1l},v_{2l})p(x_{1l},x_{2l}\mid u_l,v_{1l},v_{2l})p(y_{l},y_{1l},y_{2l}\mid x_{1l},x_{2l}).
\end{eqnarray}
 This completes the proof of Theorem 1.
 

\section{}
\label{app2}
We compute the upper bound on secrecy rate for the parallel Gaussian relay-eavesdropper channel as follows.
 \begin{align}
\label{gproof}
\max_{\{{\bf{K}_{P_\textit{l}}} \in {{\mc{K}}_{P_\textit{l}}}\}_{\textit{l}=1,\hdots, L}} &\sum_{l=1}^L I( X_{1l}, X_{2l};Y_l )-I( X_{1l}, X_{2l};Y_{2l})\notag\\
=&  \max_{\{{\bf{K}_{P_\textit{l}}} \in {{\mc{K}}_{P_\textit{l}}}\}_{\textit{l}=1,\hdots, L}}  \sum_{l=1}^L [h(Y_l ) -  h(Y_l \mid X_{1l},X_{2l}) - h(Y_{2l})+  h(Y_{2l}\mid X_{1l},X_{2l})] \notag \\
=&  \max_{\{{\bf{K}_{P_\textit{l}}} \in {{\mc{K}}_{P_\textit{l}}}\}_{\textit{l}=1,\hdots, L}}  \sum_{l=1}^L [h(Y_l) - h(Z_l) - h(Y_{2l})+h(Z_{2l})].
\end{align}
The first term in \eqref{gproof} is computed as follows.
\setlength{\arraycolsep}{0em}
\begin{align}
\label{t1}
h(Y_l)&= h( X_{1l}+\sqrt{\rho_{1l}}X_{2l}+Z_l)\notag \\
          &= \frac{1}{2}\log (2\pi e)(P_{1l}+\rho_{1l}P_{2l}+2\psi_l\sqrt{\rho_{1l}P_{1l}P_{2l}}+\sigma_{l}^2 ).
\end{align}
Similarly the second, third and fourth term in \eqref{gproof} are computed as follows.
  \begin{align}
\label{t2}
h(Z_l) &= \frac{1}{2}\log 2\pi e (\sigma_{l}^2) \\
\label{t3}
h(Y_{2l}) &= \frac{1}{2}\log (2\pi e)(P_{1l}+\rho_{2l}P_{2l}+2\psi_l\sqrt{\rho_{2l}P_{1l}P_{2l}}+\sigma_{2l}^2 )\\
\label{t4}
h(Z_{2l}) &= \frac{1}{2}\log 2\pi e (\sigma_{2l}^2).
\end{align}
 \setlength{\arraycolsep}{5pt}
Using \eqref{t1}-\eqref{t4} in \eqref{gproof} gives
\small {
\begin{align}
\label{eqf}
R_e^{\textrm{up}}= \max_{
\substack{
\sum_{l=1}^L P_{1l} \le P_1,\\ \sum_{l=1}^L P_{2l} \le P_2, \\
-1 \le {\psi_l} \le 1 \\  \text{for $l=1,\hdots,L$}}}\sum_{l=1}^L
\frac{1}{2}\log \Big(1+\frac{P_{1l}+\rho_{1l}P_{2l}+2\psi_l\sqrt{\rho_{1l}P_{1l}P_{2l}}}{\sigma_{l}^2}\Big)- \frac{1}{2}\log \Big(1+\frac{P_{1l}+\rho_{2l}P_{2l}+2\psi_l\sqrt{\rho_{2l}P_{1l}P_{2l}}}{\sigma_{2l}^2}\Big).
\end{align}
}

\bibliographystyle{IEEEtran}
\bibliography{IEEEsecrecy}

\begin{IEEEbiography}[{\includegraphics[width=1.1in,height=1.3in]{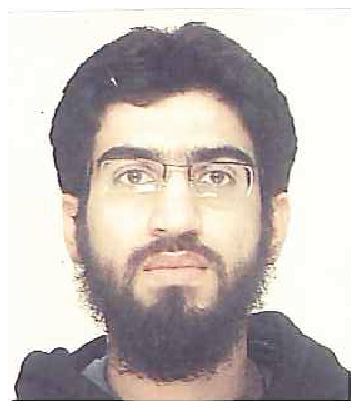}}]
{Zohaib Hassan Awan}  
 received the B.S. degree in Electronics Engineering from Ghulam Ishaq Khan Institute (GIKI), Topi, Pakistan in 2005 and the M.S. degree in Electrical Engineering with  majors in wireless systems from Royal Institute of Technology (KTH), Stockholm, Sweden in 2008. Since Jan. 2009, he has been working towards his Ph.D. degree with the ICTEAM institute, Universit\'{e} catholique de Louvain (UCL), Belgium.

His research interests include information-theoretic security, cooperative communications and communication theory.
\end{IEEEbiography}

\begin{IEEEbiography}[{\includegraphics[width=1.1in,height=1.3in]{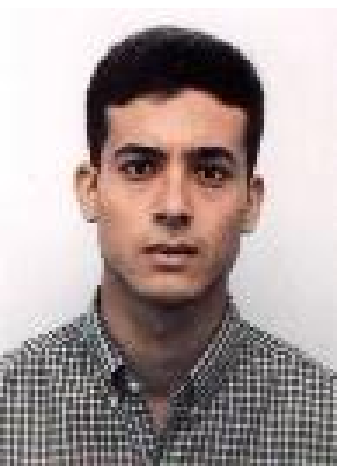}}]
{Abdellatif Zaidi}  
 received the B.S. degree in Electrical Engineering from \'{E}cole Nationale Sup\'{e}rieure de Techniques Avanc\'{e}s, ENSTA ParisTech, France
in 2002 and the M. Sc. and Ph.D. degrees in Electrical Engineering from \'{E}cole Nationale Sup\'{e}rieure des T\'{e}l\'{e}communications, TELECOM ParisTech, Paris, France in 2002 and 2005, respectively.

From December 2002 to December 2005, he was with the Communications and Electronics Dept., TELECOM ParisTech, Paris, France and the Signals and
Systems Lab., CNRS/Sup\'{e}lec, France pursuing his PhD degree. From May 2006 to September 2010, he was at \'{E}cole Polytechnique de Louvain, Universit\'{e} catholique de Louvain, Belgium, working as a research assistant. Dr. Zaidi was "Research Visitor" at the University of Notre Dame, Indiana, USA,
during fall 2007 and Spring 2008. He is now, an assistant professor at Universit\'e Paris-Est Marne-la-Vall\'ee, France.

His research interests cover a broad range of topics from signal processing for communication and multi-user information theory. Of particular
interest are the problems of coding for side-informed channels, secure communication, coding and interference mitigation in multi-user channels,
and relaying problems and cooperative communication with application to sensor networking and ad-hoc wireless networks.
\end{IEEEbiography}

\begin{IEEEbiography}[{\includegraphics[width=1.1in,height=1.3in]{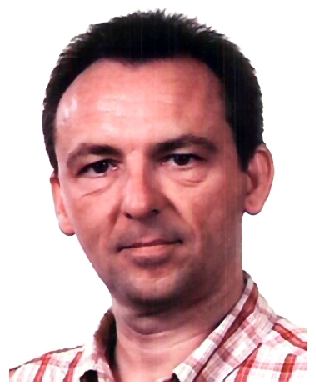}}]
{Luc Vandendorpe} (M'93-SM'99-F'06)
 was born in Mouscron, Belgium, in 1962. He received the Electrical Engineering degree (summa cum laude) and the Ph. D. degree from the Universit\'{e} catholique de Louvain (UCL) Louvain-la-Neuve, Belgium in 1985 and 1991 respectively. Since 1985, L. Vandendorpe is with the Communications and Remote Sensing Laboratory of UCL where he first worked in the field of bit rate reduction techniques for video coding. In 1992, he was a Visiting Scientist and Research Fellow at the Telecommunications and Traffic Control Systems Group of the Delft Technical University, Netherlands, where he worked on Spread Spectrum Techniques for Personal Communications Systems. From October 1992 to August 1997, L. Vandendorpe was Senior Research Associate of the Belgian NSF at UCL. Presently, he is Full Professor and Head of the Institute for Information and Communication Technologies, Electronics and Applied Mathematics of UCL.

His current interest is in digital communication systems and more precisely resource allocation for OFDM(A) based multicell systems, MIMO and distributed MIMO, sensor networks, turbo-based communications systems, physical layer security and UWB based positioning.

In 1990, he was co-recipient of the Biennal Alcatel-Bell Award from the Belgian NSF for a contribution in the field of image coding. In 2000 he was co-recipient (with J. Louveaux and F. Deryck) of the Biennal Siemens Award from the Belgian NSF for a contribution about filter bank based multicarrier transmission. In 2004 he was co-winner (with J. Czyz) of the Face Authentication Competition, FAC 2004. L. Vandendorpe is or has been TPC member for numerous IEEE conferences (VTC Fall, Globecom Communications Theory Symposium, SPAWC, ICC) and for the Turbo Symposium. He was co-technical chair (with P. Duhamel) for IEEE ICASSP 2006.

He was an editor of the IEEE Trans. on Communications for Synchronization and Equalization between 2000 and 2002, associate editor of the IEEE Trans. on Wireless Communications between 2003 and 2005, and associate editor of the IEEE Trans. on Signal Processing between 2004 and 2006. He was chair of the IEEE Benelux joint chapter on Communications and Vehicular Technology between 1999 and 2003. He was an elected member of the Signal Processing for Communications committee between 2000 and 2005, and between 2009 and 2011, and an elected member of the Sensor Array and Multichannel Signal Processing committee of the Signal Processing Society between 2006 and 2008. Currently, he is the Editor in Chief for the Eurasip Journal on Wireless Communications and Networking. L. Vandendorpe is a Fellow of the IEEE.
\end{IEEEbiography}

\end{document}